\documentclass{aa}
\usepackage{amsmath,amssymb}
\usepackage{epsf}

\makeatletter
\def\thebibliography#1{\section*{REFERENCES}
 \addcontentsline{toc}{section}{REFERENCES}
 \list{}{\labelwidth\z@
         \leftmargin 1.5em
	 \itemsep \z@
	 \itemindent-\leftmargin}
 \small\raggedright
 \parindent\z@
 \parskip\z@ plus .1pt\relax
 \def\newblock{\hskip .11em plus .33em minus .07em}
 \sloppy\clubpenalty4000\widowpenalty4000
 \sfcode`\.=1000\relax
}

\def\@biblabel#1{}
\def\@bcite#1#2{(#1\if@tempswa , #2\fi)}
\def\@pcite#1#2{#1\if@tempswa , #2\fi}
\def\@citefmta#1#2{#1 (#2)}
\def\@citefmtb#1#2{#1 #2}
\let\citefmt=\@citefmta

\def\@citex[#1]#2{\if@filesw\immediate\write\@auxout{\string\citation{#2}}\fi
  \def\@citea{}\@cite{\@for\@citeb:=#2\do
    {\@citea\def\@citea{;\penalty\@m\ }\@ifundefined
    {b@\@citeb}{{\bf ?}\@warning
{Citation `\@citeb' on page \thepage \space undefined}}%
{\csname b@\@citeb\endcsname}}}{#1}}

\def\cite{\@ifnextchar [{\let\citefmt=\@citefmtb
                          \let\@cite=\@bcite\@tempswatrue \@citex}
                        {\let\citefmt=\@citefmtb
                          \let\@cite=\@bcite\@tempswafalse \@citex[]}}
\def\pcite{\@ifnextchar [{\let\citefmt=\@citefmtb
                          \let\@cite=\@pcite\@tempswatrue\@citex}
                        {\let\citefmt=\@citefmtb
                          \let\@cite=\@pcite\@tempswafalse\@citex[]}}
\def\scite{\@ifnextchar [{\let\citefmt=\@citefmta
                          \let\@cite=\@pcite\@tempswatrue\@citex}
                        {\let\citefmt=\@citefmta
                          \let\@cite=\@pcite\@tempswafalse\@citex[]}}
\makeatother

\def\uminus{\mathbin{\cup\kern-4.5pt\mbox{-\kern-3.08331pt}\kern4.5pt}}

\def\hMpc{\ifmmode{h^{-1}{\rm Mpc}}\else{$h^{-1}$Mpc}\fi}
\def\hGpc{\ifmmode{h^{-1}{\rm Gpc}}\else{$h^{-1}$Gpc}\fi}

\def\d{{\rm d}}
\def\e{{\rm e}}
\def\P{{\rm P}}
\def\bx{{\mathbf{x}}}
\def\by{{\mathbf{y}}}
\def\bz{{\mathbf{z}}}
\def\cA{{\mathcal{A}}}
\def\cB{{\mathcal{B}}}
\def\RR{{\mathbb{R}}}

\begin{document}

\thesaurus{02(12.12.1, 12.03.4, 12.03.3, 11.03.1)}

\title{Fluctuations in the IRAS 1.2 Jy Catalogue}

\author{Martin Kerscher\inst{1} \and Jens Schmalzing\inst{1,2} \and 
Thomas Buchert\inst{1} \and Herbert Wagner\inst{1}\thanks{emails: kerscher, jens, buchert, wagner@stat.physik.uni-muenchen.de}}

\institute{Ludwig--Maximilians--Universit\"at, Theresienstra{\ss}e 37, 
80333 M\"unchen, Germany
\and
Max--Planck--Institut f\"ur Astrophysik,
Karl--Schwarzschild--Stra{\ss}e 1, 85740 Garching, Germany}

\offprints{T.\ Buchert}

\date{Received 6 October 1997 / Accepted 23 December 1997}

\maketitle

\begin{abstract}
An analysis of the IRAS 1.2~Jy redshift catalogue with emphasis on the
separate examination of    northern and southern parts   (in  galactic
coordinates) is  performed using a    complete set  of   morphological
descriptors  (Minkowski functionals), nearest neighbour  distributions
and  the variance of the galaxy  counts. We find large fluctuations in
the clustering  properties as seen  in a large  difference between the
northern   and  southern  parts    of  the   catalogue  on   scales of
100\hMpc. These fluctuations  remain discernible even  on the scale of
200\hMpc. We  also identify  sparse   sampling as  a  major  source of
``apparent homogenization''.  Tests on observational selection effects
concerning luminosity, colour  and  redshift--space distortion support
the significance of these large--scale fluctuations.
\keywords{Cosmology: large--scale structure of  the Universe -- theory
-- observations -- Galaxies: clustering}
\end{abstract}

\section{Introduction}

The   spatial distribution of  luminous   matter seen in the available
galaxy redshift catalogues    conveys a  complex  picture  of   voids,
isolated  galaxies, clusters   and  superclusters, with   the  largest
filamentary features  extending on scales  comparable with the size of
the surveys.   Since these findings  are expected to provide important
constraints for  theoretical  models of  structure evolution,  a prime
task is to  extract the morphological  characteristics of the observed
patterns by appropriate quantitative measures.

In  the    present paper  we    consider  the IRAS   1.2~Jy  catalogue
{}\cite{fisher:irasdata}.   This infrared--selected survey consists of
two data   sets for  the   northern and   southern caps  (in  galactic
coordinates) separated by  a 10 degree wide zone  of  avoidance and is
therefore nearly all--sky.  We analyze the two subsamples individually
with Minkowski functionals  {}\cite{mecke:robust} in order to quantify
the     global  geometrical  and   topological    features  of spatial
patterns. In addition, we apply a novel  method of cluster diagnostics
{}\cite{vanlieshout:j}  combining local and
global statistical    descriptors,    and  the      more  conventional
count--in--cell method.

Practically,   when analyzing  galaxy surveys  one   has to deal  with
various technical   problems.   As a rule,   three--dimensional galaxy
catalogues are  constructed from  redshift  surveys which  suffer from
distortions due  to  peculiar motions.  To  date,  there is no  way of
tackling     this   problem    without   making    strong  assumptions
(e.g.~linearized  equations of  motion), and therefore  we conduct our
analysis in  redshift  space and investigate  the differences  to real
space   by  comparing  with     results  of   $N$--body   simulations.
Furthermore, surveys are usually flux limited. Therefore, the observed
number density of  the galaxies decreases  with distance.  For samples
with inhomogeneous     mean  number  density    the  methods employing
morphological estimators   are  usually dominated  by thinning  at the
fringes. One remedy  is to focus on volume  limited subsamples, at the
cost of losing a  substantial fraction of galaxies.  Nevertheless,  we
show that Minkowski functionals provide significant results even for a
small number of galaxies.  Finally,  all surveys are spatially limited
both in directions as well as  in depth, where volume limitation leads
to a well--defined spherical boundary. Hence, we  have to take care of
boundary  constraints.  In Appendix~A   we  discuss these finite  size
effects  on the  estimators for  the  Minkowski  functionals, for  the
nearest--neighbour   distribution,  and for   the  fluctuations of the
galaxy counts.

As a result we detect a  significant morphological segregation between
the  northern  and the southern  part   of the IRAS  catalogue.  These
findings are also supported by our further results on the variances of
counts in cells, done separately for the north and the south.

\section{Methods of morphological analysis}

We consider a set $X=\{\bx_i\}_{i=1}^N$ of $N$ points $\bx_i\in\RR^3$
given by the redshift--space coordinates of the galaxies in the IRAS
1.2~Jy catalogue {}\cite{fisher:irasdata}.

\subsection{Minkowski functionals}

A   computationally convenient and  robust method  for quantifying the
morphology of spatial patterns is  offered by Minkowski functionals as
they were introduced   into cosmology  by  {}\scite{mecke:robust}. For
this purpose we decorate each point $\bx_i$ with a ball $\cB_r(\bx_i)$
of  radius~$r$      and      then    consider   the        union   set
$\cA_N(r)=\bigcup_{i=0}^N\cB_r(\bx_i)$.   {}\scite{hadwiger:vorlesung}
proved that  in   three dimensions   the   four Minkowski  functionals
$M_{\mu=0,1,2,3}(\cA_N(r))$   give  a     complete       morphological
characterization  of a  body  $\cA_N(r)$. The  interpretation of these
functionals in terms  of  geometrical  and topological quantities   is
given in Table~(\ref{table:mingeom}).

\begin{table}
\caption{\label{table:mingeom}
Minkowski functionals in three--dimensional space expressed in terms
of more familiar geometric quantities.}
\vspace{0.5cm}
\begin{flushleft}
\begin{tabular}{lllll}
\hline\noalign{\smallskip}
       & geometric quantity      & $\mu$ & $M_\mu$      & $\Phi_\mu$     \\
\noalign{\smallskip}
\hline\noalign{\smallskip}
$V$    & volume                  & 0 	 & $V$ & $V/(\tfrac{4\pi}{3}r^3N)$ \\
$A$    & surface                 & 1 	 & $A/8$        & $A/(4\pi r^2N)$  \\
$H$    & integral mean curvature & 2 	 & $H/(2\pi^2)$   & $H/(4\pi rN)$    \\
$\chi$ & Euler characteristic    & 3 	 & $3\chi/(4\pi)$ & $\chi/N$       \\
\noalign{\smallskip}
\hline\noalign{\smallskip}
\end{tabular}
\end{flushleft}
\end{table}

By normalizing with the functional $M_\mu(\cB_r)$  of a single ball we
can  introduce  normalized,    dimensionless  Minkowski    functionals
$\Phi_\mu(\cA_N(r))$,
\begin{equation}\label{eq:Phi-def}
\Phi_\mu(\cA_N(r)) := \frac{M_\mu(\cA_N(r))}{NM_\mu(\cB_r)}.
\end{equation}
In the  case   of a Poisson process   the  exact mean  values  of  the
functionals can be  calculated analytically  {}\cite{mecke:euler}. For
decorating spheres with radius $r$ one obtains:
\begin{equation}\label{eq:Poisson}
\begin{split}
\Phi_0^\P &=\left(1-\e^{-\eta}\right)~\eta^{-1}, \\
\Phi_1^\P &=\e^{-\eta}, \\
\Phi_2^\P &=\e^{-\eta}~(1-\tfrac{3\pi^2}{32}\eta), \\
\Phi_3^\P &=\e^{-\eta}~(1-3\eta+\tfrac{3\pi^2}{32}\eta^2), 
\end{split}
\end{equation}
with                  the               dimensionless        parameter
$\eta=\overline{n}M_0(\cB_r)=\overline{n}\      4\pi  r^3/3$,    where
$\overline{n}$ denotes the mean number density.

For   $\mu\ge1$   the  measures   $\Phi_{\mu}(\cA_N(r))$  contain  the
exponentially decreasing factor $\e^{-\eta(r)}$. It  can be shown that
this exponential decrease with increasing (diagnostic) radius $r$ also
arises for more  general cluster  processes.  Therefore we employ  the
reduction
\begin{equation}\label{eq:phi-def}
 \phi_\mu(\cA_N(r)) =
 \frac{\Phi_\mu(\cA_N(r))}{\Phi_1^\P(\cA_N(r))}, \quad \mu
 \ge 1,
\end{equation}
and thereby remove the exponential decay and enhance the visibility of
differences in the displays shown below.

\subsection{Nearest neighbour statistics}
\label{sec:nextneighbour}
The nearest neighbour distribution $G(r)$   is a standard tool in  the
analysis of point processes {}\cite{ripley:spatial}  and is defined as
the {\em distribution of  distances $r$ of a point  of the process  to
the nearest  other  point of  the  process}.  The function  $G(r)$  is
related to conditional correlation functions {}\cite{white:hierarchy}.
Another common statistical descriptor  is   the empty space   function
$F(r)$, the  {\em distribution  of the distances  $r$ of  an arbitrary
point to the   nearest point  of the  process},  and is  equal to  the
expected fraction of volume occupied by  points which are less distant
than  $r$  from the next point   of the process.  Therefore, $F(r)$ is
equal to  the  volume density    of  $M_0(\cA_N(r))$; hence   $1-F(r)$
coincides   with  the      void   probability   function      $P_0(r)$
{}\cite{white:hierarchy}.  Alternatively,  one may interpret $F(r)$ as
the unconditional  and $G(r)$ as the   conditional distribution of the
nearest  neighbour    distance.   Recently,    {}\scite{vanlieshout:j}
advocated to use the ratio
\begin{equation}
J(r) := \frac{1-G(r)}{1-F(r)}
\end{equation}
as a probe for clustering of a point process. In the case of a Poisson
process,
\begin{equation}
G^\P(r) = 1 - \exp(- \eta) = F^\P(r) ,
\end{equation}
and thus $J^\P(r) = 1$. For a  process with enhanced clumping, we have
$J(r) <  1$, whereas regular structures are  indicated by  $J(r) > 1$.

\subsection{Fluctuations of the galaxy counts}

Another frequently  used method for exploring  galaxy catalogues is to
consider  the   fluctuations  of galaxy  counts,    in particular, the
variance of counts in cells in excess of Poisson,
\begin{equation}\label{eq:sigma-def1}
\left\langle(N_i-\overline{n}V)^2\right\rangle = 
\overline{n}V+\overline{n}^2 V^2\sigma^2 ,
\end{equation}
where $N_i$ is the  number of galaxies in  a  volume V, in our  case a
ball $\cB_r$  with radius $r$; the  averaging is performed over random
positions of the  balls.  For a Poisson   process, $\sigma^2 =   0$ by
definition. For stationary point processes  the variance $\sigma^2$ is
related to the two--point correlation function $\xi(r)$,
\begin{equation}
\sigma^2 = \frac{1}{V^2}\int_V\int_V\d^3x\d^3y\,\xi(|x-y|).
\end{equation}
After      averaging a       power--law    correlation        function
$\xi(r)=(r_0/r)^\gamma$    over        balls  $\cB_r$    one   obtains
{}\cite{peebles:principles}:
\begin{equation}\label{eq:sigma-powerlaw}
\sigma^2(r) = \Big(\frac{r_0}{2r} \Big)^\gamma 
              \frac{1}{(1-\gamma/3)(1-\gamma/4)(1-\gamma/6)}.
\end{equation}
In Appendix~B we discuss the  dependence of the variance $\sigma^2$ on
the mean number  density and we point out  how this has to be modified
for fractal sets.

\section{Results}
We  now  apply  the methods   introduced above to   explore a redshift
catalogue of 5313 IRAS selected galaxies  with limiting flux of 1.2~Jy
{}\cite{fisher:irasdata}.

\subsection{Volume limited samples with 100\hMpc\ depth}
\begin{figure}
\begin{center} 
\epsfxsize=7cm
\begin{minipage}{\epsfxsize}\epsffile{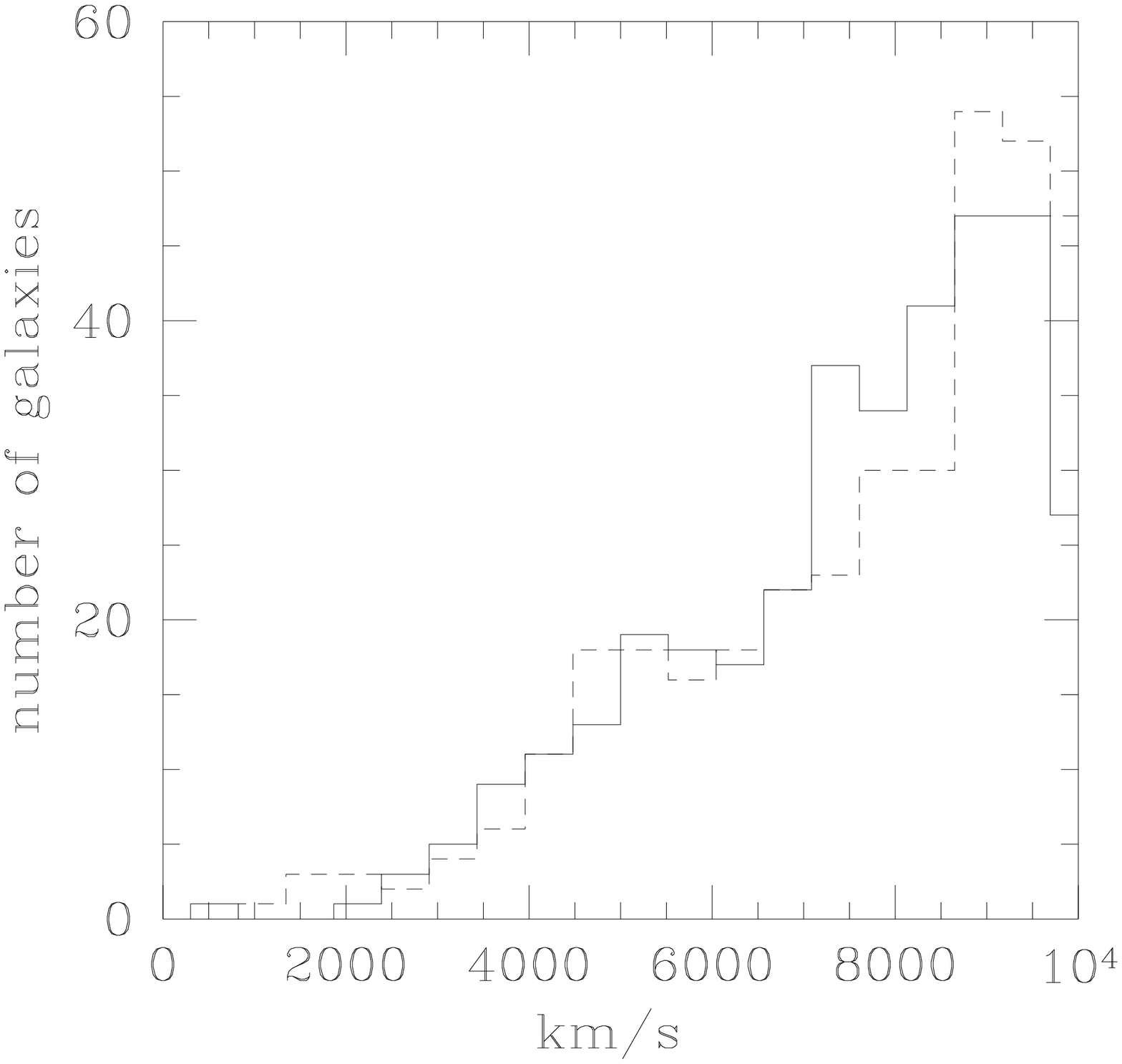}\end{minipage} 
\epsfxsize=7cm
\begin{minipage}{\epsfxsize}\epsffile{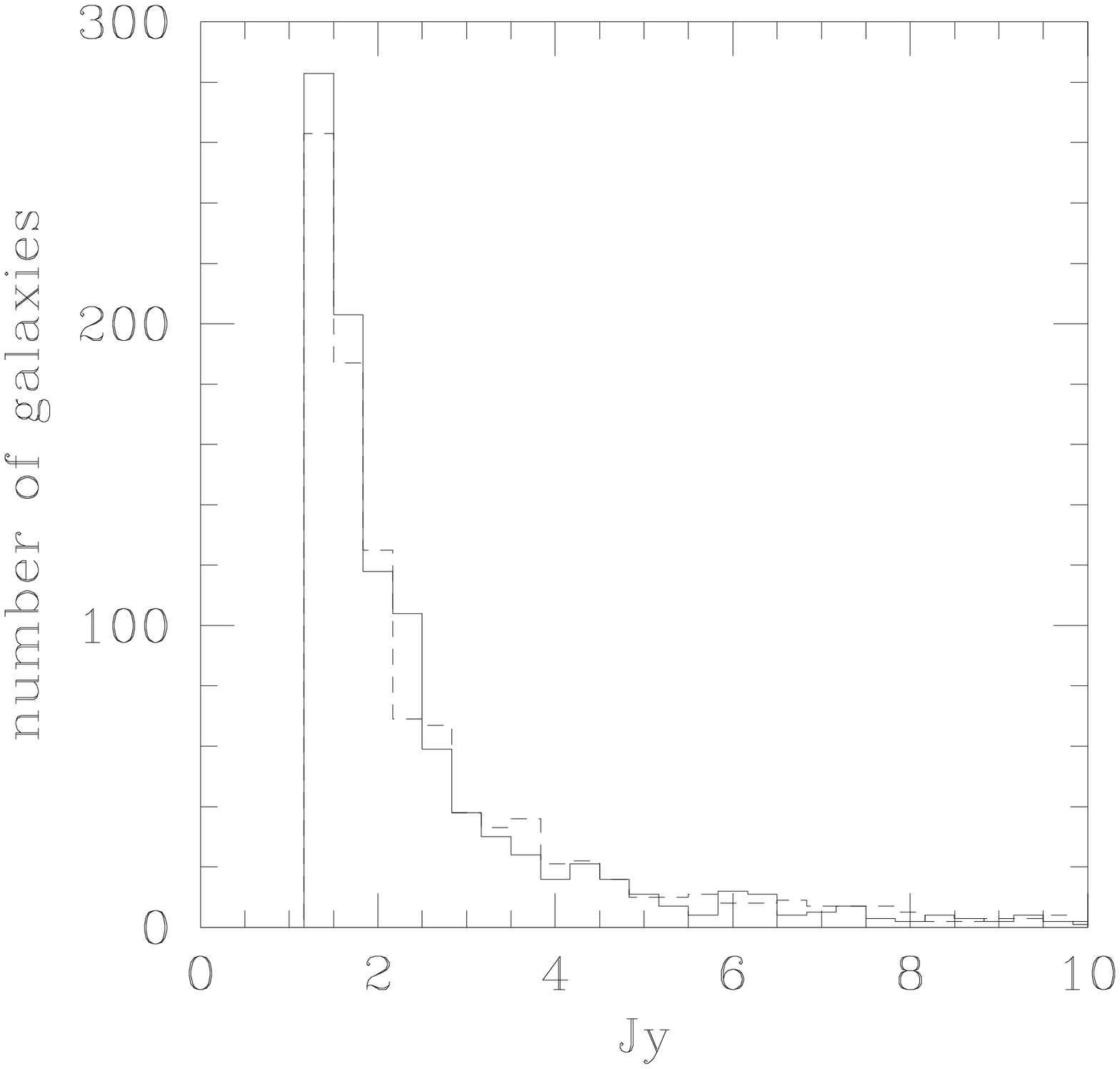}\end{minipage} 
\end{center}
\caption{\label{fig:jyv10-hist}   In the   top  panel   we  show   the
histograms of the redshift distribution of the galaxies for the volume
limited sample with 100\hMpc\  depth. In the bottom  panel we show the
histograms for the flux distribution. The northern  part is shown as a
solid line, and the southern part as a dashed line.}
\end{figure}

A volume limited  sample  of 100\hMpc\  depth   (with $h$ defined   by
$H_0=100h\  {\rm  km}\ {\rm  s}^{-1}\  {\rm  Mpc}^{-1}$) contains  352
galaxies in the northern part, and  358 galaxies in the southern part.
In  Fig.~{}\ref{fig:jyv10-hist} we see   that  the redshift and   flux
distributions  of  the galaxies  do   not indicate any  peculiarity or
problem with  the  sample.  As far  as the  number  density, i.e.\ the
first moment of the galaxy distribution is  concerned, the sample does
not reveal differences  between north and  south.  However, we want to
assess the  clustering properties of  the data and, above  all, tackle
the  question  whether the   southern and  northern  parts  differ  or
not. More   refined measurement of   the  galaxy distribution  need to
incorporate     at      least     the      second      moment     (see
Sect.~{}\ref{sec:nn_sigma_v10}, {}\ref{sec:nn_sigma_v20}).  A complete
characterization   of morphology,  depending on   all  moments of  the
distribution, is even more desirable.  It is provided by the Minkowski
functionals.

\subsubsection{Minkowski functionals}

\begin{figure}
\begin{center} 
\epsfxsize=7cm
\begin{minipage}{\epsfxsize}\epsffile{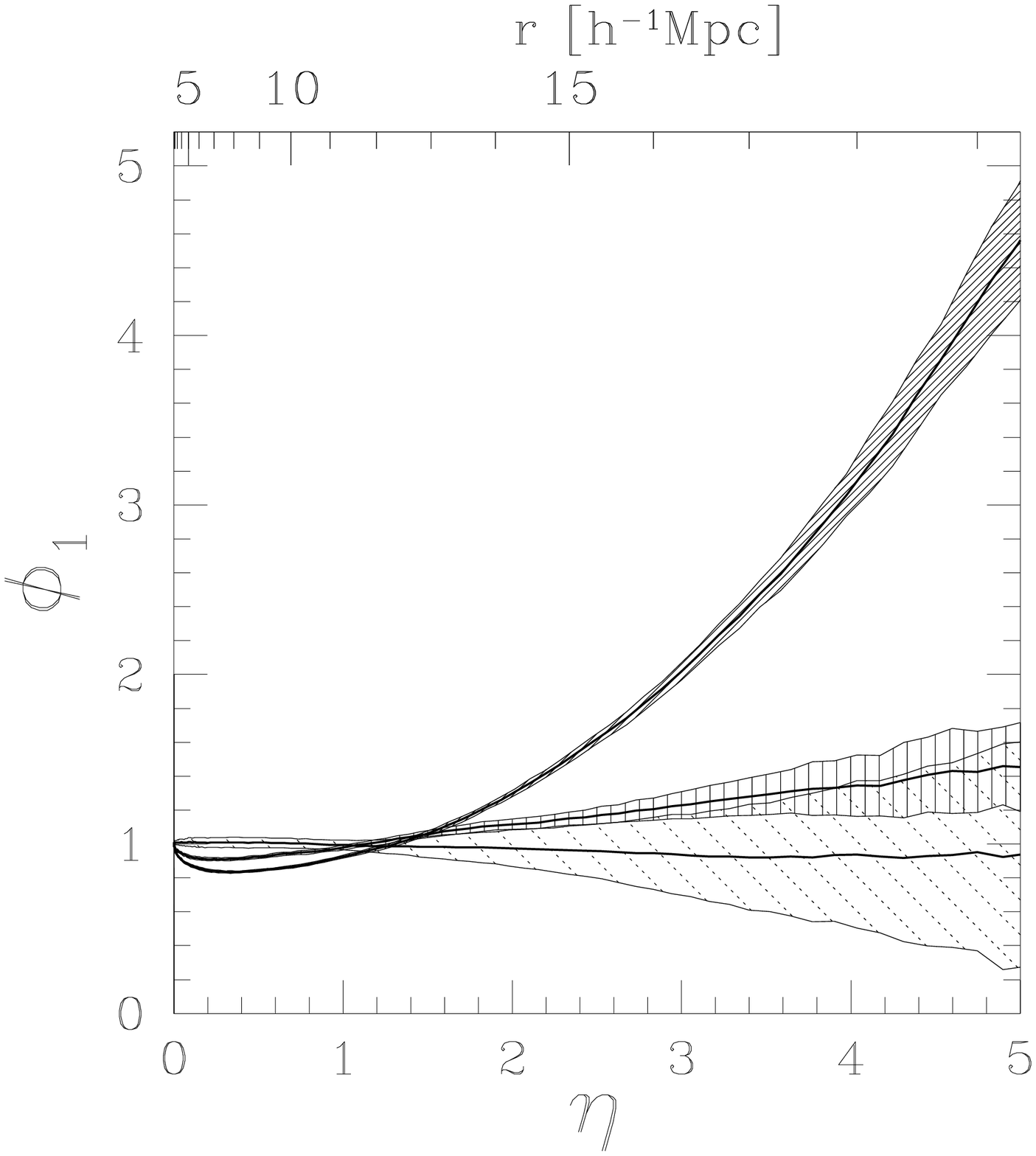}\end{minipage}
\epsfxsize=7cm
\begin{minipage}{\epsfxsize}\epsffile{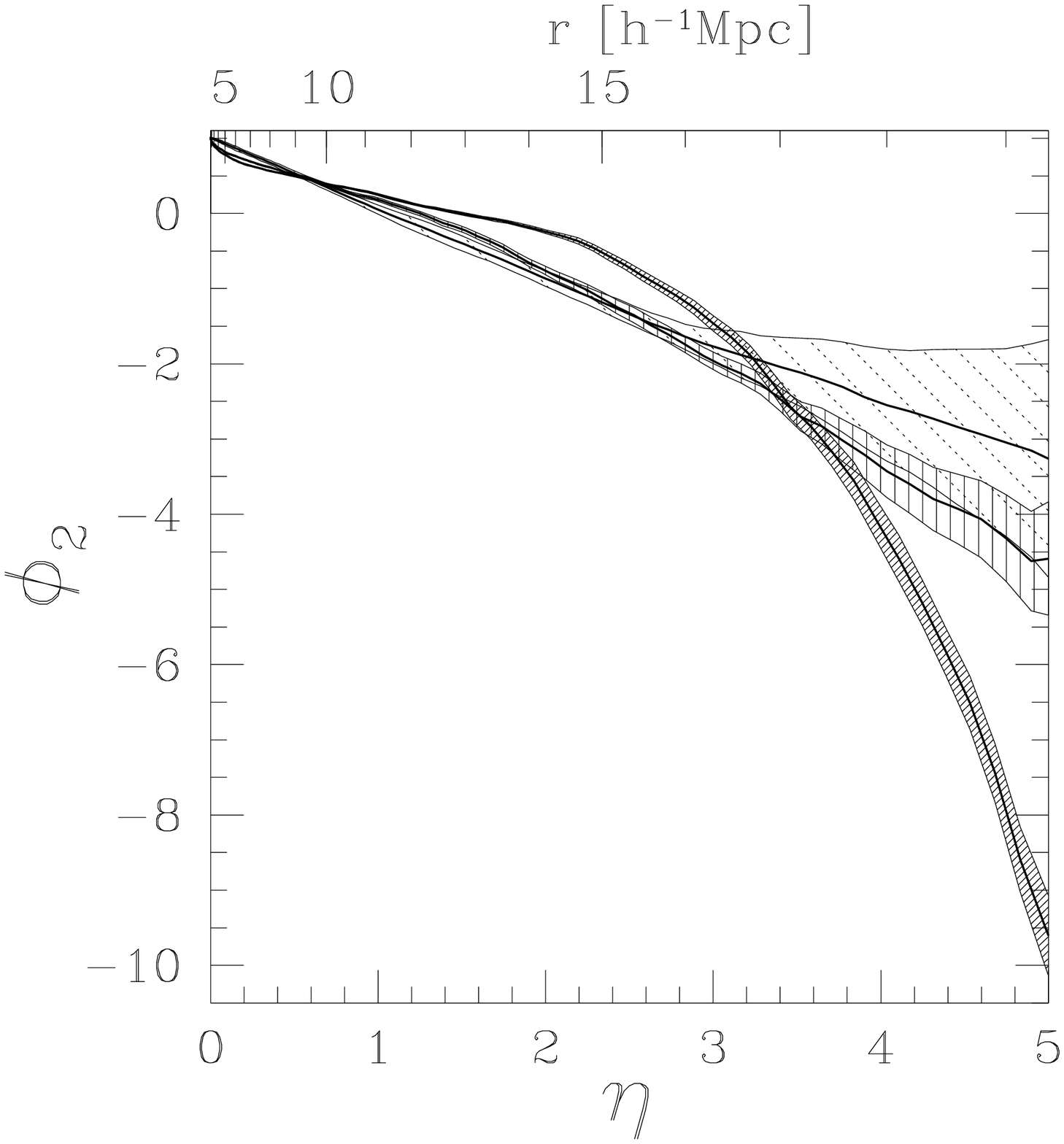}\end{minipage}
\epsfxsize=7cm
\begin{minipage}{\epsfxsize}\epsffile{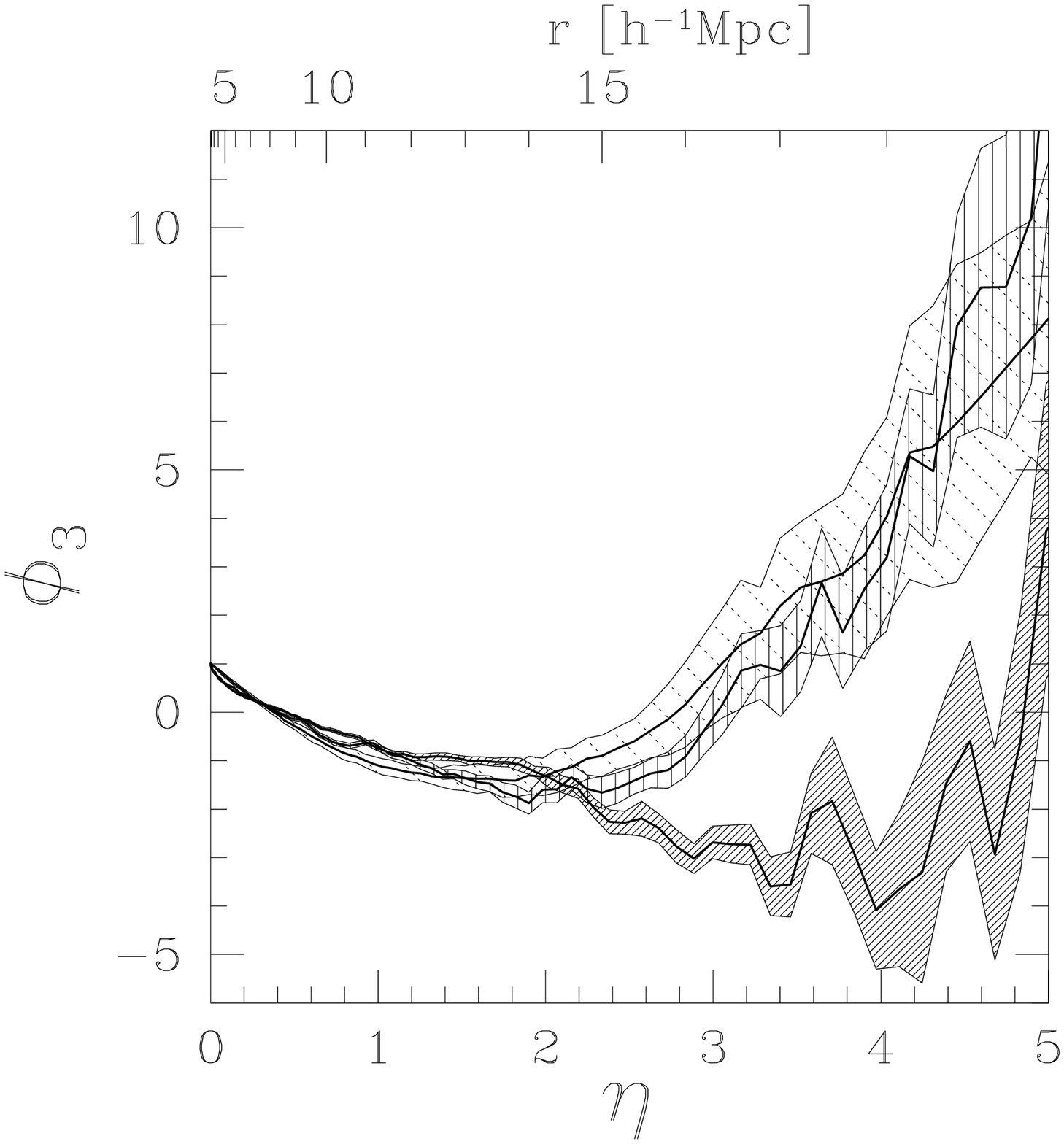}\end{minipage}
\end{center}
\caption{\label{fig:minjyv10} Minkowski  functionals $\phi_\mu$  of  a
volume limited sample with  100\hMpc\   depth; the dark  shaded  areas
represent the southern part, the  medium shaded the northern part, and
the dotted a Poisson process with the same  number density. The shaded
areas are the $1\sigma$ errors as explained in the text.}
\end{figure}

In  Fig.~\ref{fig:minjyv10}  we show  the    values of  the  Minkowski
functionals of  the  southern and  northern  parts, together  with the
values from a Poisson process with the same number density (the volume
density  is  shown  in Fig.~\ref{fig:FGJjyv10}).  The   errors for the
Poisson    process   were    calculated   using   twenty     different
realizations. This already  gives a  stable  estimate of  the ensemble
variance.  To estimate the error for  the catalogue data we calculated
the Minkowski functionals of twenty  subsamples containing 90\% of the
galaxies, randomly chosen from the volume limited subsample.

In both parts of  the 1.2~Jy catalogue the  clustering of  galaxies on
scales up  to 10\hMpc\  is  clearly stronger than  in   the case  of a
Poisson process, as inferred from  the lower values of the functionals
for the surface area, $\phi_1$, the integral mean curvature, $\phi_2$,
and the Euler characteristic,   $\phi_3$.  Moreover, the  northern and
southern parts differ significantly, with the northern part being less
clumpy. The  most conspicuous features  are the enhanced  surface area
$\phi_1$ in the southern  part on scales from  12 to 20\hMpc\ and  the
kink in the integral mean curvature $\phi_2$ at 14\hMpc. This behavior
indicates that dense substructures in the southern  part are filled up
at this scale (i.e.~the  balls in these substructures  overlap without
leaving holes), and is presumably the signature of the Perseus--Pisces
supercluster   (for       a      more     detailed     analysis    see
{}\pcite{kerscher:significance}).  On scales    from 15  to   20\hMpc,
$\phi_2$,  the  value  of the integral   mean  curvature is  negative,
indicating  concave structures.    In  the  southern part   the  Euler
characteristic is  still negative   in   this range;   therefore,  the
structure  is  dominated  by  interconnected  tunnels (giving  rise to
negative contributions to  the Euler  characteristic)  rather than  by
completely   enclosed  voids     (these  would   result  in   positive
contributions to the Euler characteristic).  A similar feature is seen
in    the      Minkowski   functionals    of      Abell/ACO   clusters
{}\cite{kerscher:abell}.

\subsubsection{Error estimates}
\label{sec:errors}
To get an impression how the error from  subsampling is related to the
intrinsic variance of an ensemble we looked at fixed realizations of a
Poisson process within the  sample geometry  and calculated the  error
via subsampling using again 90\%  of the points.  This error turns out
to  be  two  times smaller than  the   ensemble error  calculated over
different realizations of a Poisson process.

As  for a further error estimate  we randomize the redshifts using the
quoted redshift  errors as the standard  deviation (and using the mean
redshift error, if   none is quoted).   These  errors are found  to be
approximately two  times smaller than the   errors from subsampling as
shown in Fig.~\ref{fig:minjyv10}.   Even  if we  increase  the  quoted
redshift error by  a factor of five  the errors in the functionals are
of the same order as determined from subsampling.

In using  a window as shown  in Fig.~\ref{fig:window} we take  care of
the zone  of avoidance.  There  are additional  holes in the  redshift
catalogue  due to a lack  of  sky coverage  or  confusion in the point
source catalogue. In the  northern part these  holes account for 3.2\%
of the  catalogue, in  the southern  part  they account for  4.5\%. To
estimate the influence of these  regions on the morphological measures
we throw Poisson distributed points with the  same number density into
these regions.  The  additional  error introduced from    these random
points is smaller  than the error from  randomizing  the redshifts and
therefore, much smaller than the errors from subsampling. Moreover, no
systematic effect is seen, the curves overlap completely.

\subsubsection{Selection effects and the effect of sparse sampling}
\label{sec:selection}

\begin{figure}
\begin{center} 
\epsfxsize=7cm
\begin{minipage}{\epsfxsize}\epsffile{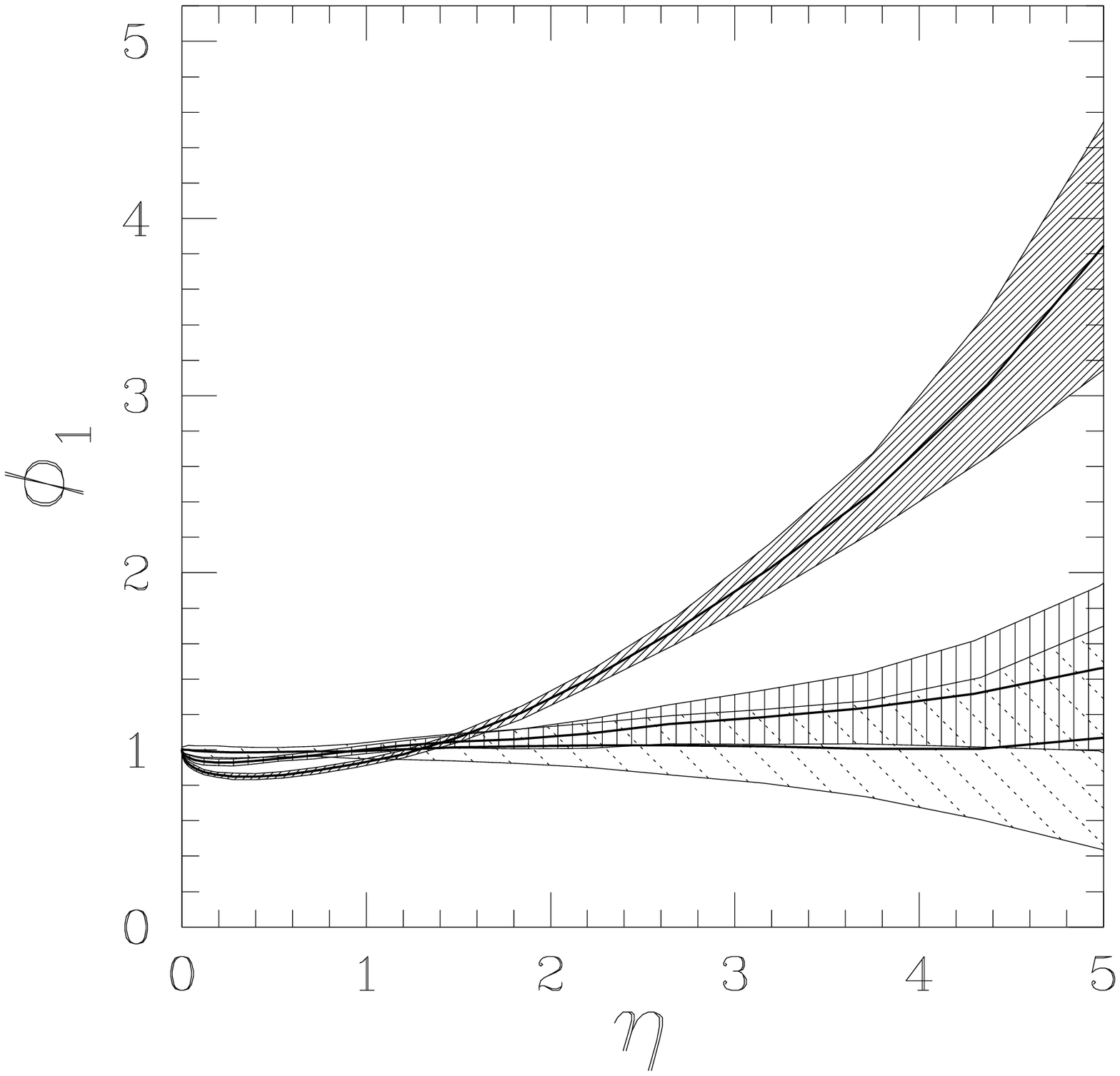}\end{minipage}
\epsfxsize=7cm
\begin{minipage}{\epsfxsize}\epsffile{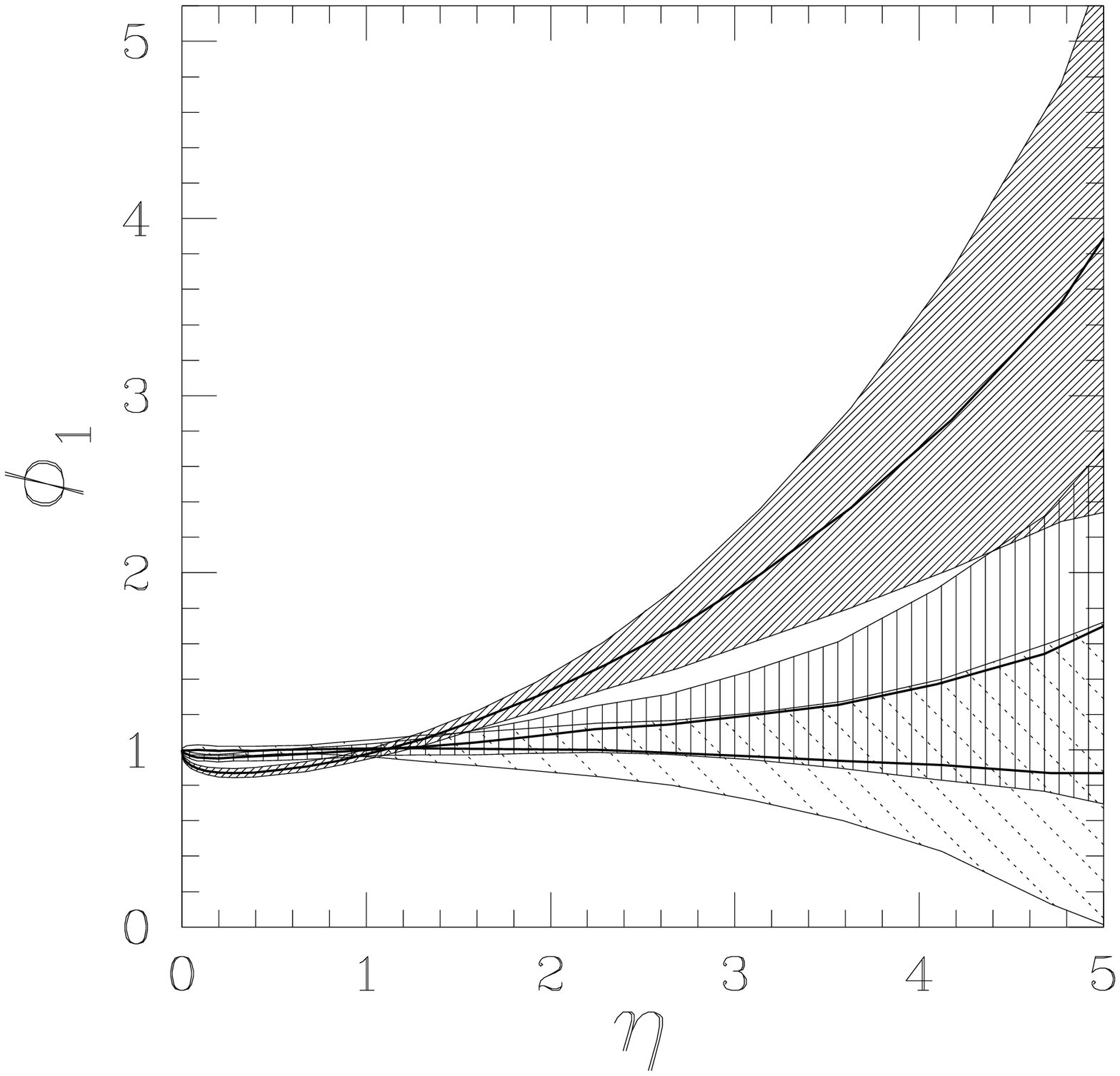}\end{minipage}
\epsfxsize=7cm
\begin{minipage}{\epsfxsize}\epsffile{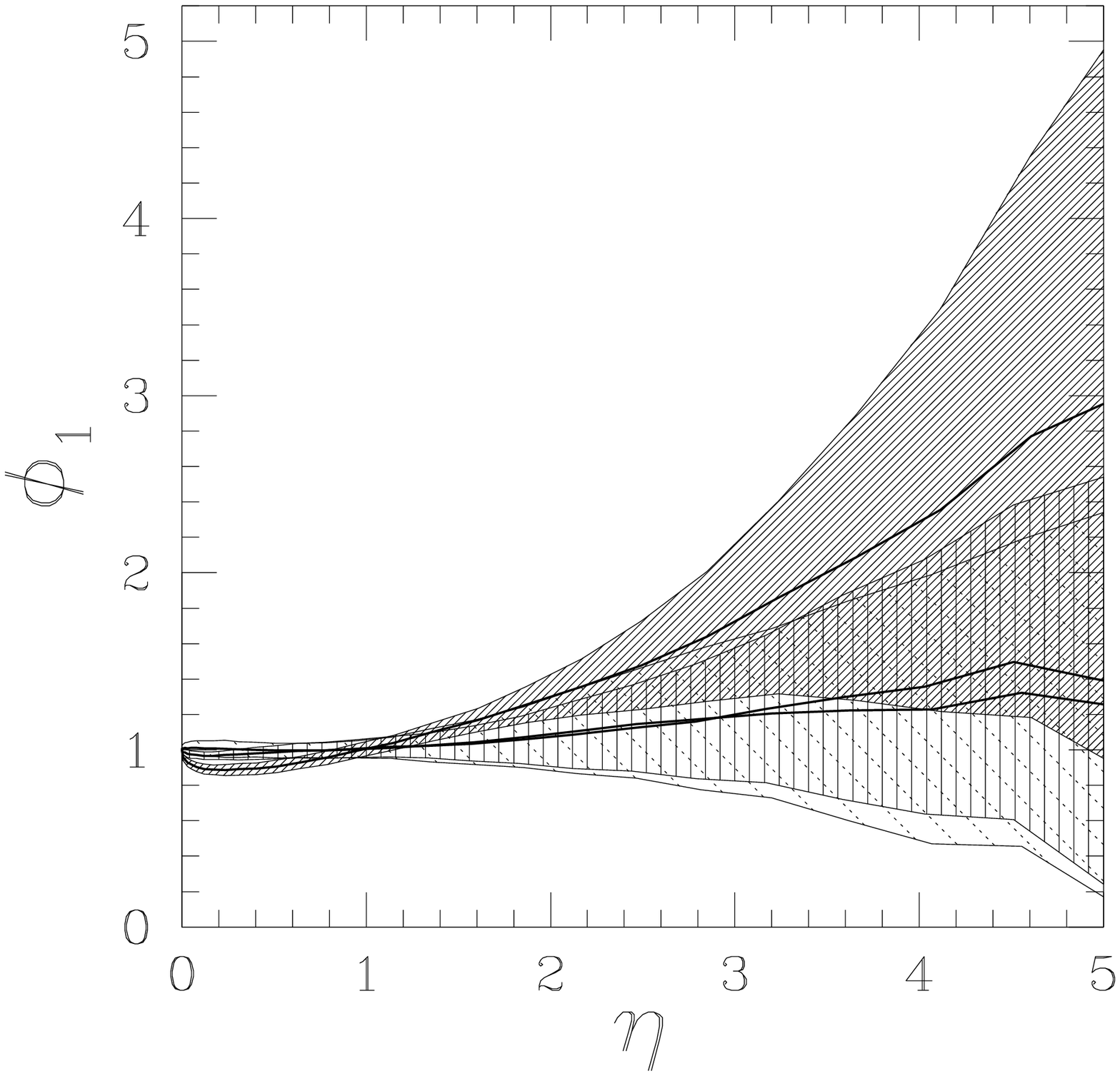}\end{minipage}
\end{center}
\caption{\label{fig:dilute} The surface functional  $\phi_1(\cA_N(r))$
of a  volume limited  sample  with 100\hMpc\ depth for  randomly drawn
subsamples with  70\% (top   panel), 50\%  (middle  panel), and   30\%
(bottom panel) galaxies.     Again, dark shaded areas   represent  the
southern part, the medium shaded  the northern part,  and the dotted a
Poisson process with the same number density.}
\end{figure}

Several selection  effects might enter   into the construction of  the
catalogue. Therefore, we draw subsamples selected according to special
features   like  ``colours'',   higher limiting  flux,  and  different
boundaries.

Before  listing these tests we address  the important issue of `sparse
sampling'. This issue is especially relevant for the interpretation of
our tests, since all  these selected samples incorporate less galaxies
than    the   volume limited    sample   with   100\hMpc\    depth. In
Fig.~\ref{fig:dilute} we show how  sparse sampling affects the surface
functionals $\phi_1(\cA_N(r))$ of   the  volume limited  samples  with
100\hMpc\ by only taking a fraction  of the galaxies into account (the
other  functionals behave  similarly).   By  reducing the  number   of
galaxies  the  error increases and the  mean  value tends  towards the
value for a Poisson process. A similar tendency is seen in the sparser
volume       limited     sample    with     depth   200\hMpc\     (see
Fig.~\ref{fig:minjyv20}).

In view of these findings the following tests were performed:
\begin{itemize}
\item
We  calculated  the Minkowski  functionals of a  volume limited sample
with   100\hMpc\ depth but now  with  limiting  flux  equal to 2.0~Jy.
Although the noise  increases  for larger radii, since  fewer galaxies
enter,  the above   mentioned  features  and the difference    between
northern and southern parts are clearly seen.
\item
We   selected     ``hot''   galaxies,       with     a    flux   ratio
$f_{100}/f_{60}\leq{}1.5$, ``warm'' ones with $1.5\leq{}f_{100}/f_{60}
\leq 3$ and ``cold'' galaxies  with $f_{100}/f_{60} \geq 3$; $f_{100}$
and $f_{60}$ denote the flux at $100\mu$ and $60\mu$, respectively. We
calculated  the Minkowski  functionals of volume  limited samples with
100\hMpc\ taken  from the ``hot''  (106  in the north   and 116 in the
south) and ``warm'' (239 in the  north and 227  in the south) galaxies
(only 7 (north) and 15 (south) galaxies are ``cold''). Again the error
increases, but the distinct features between north and south are still
discernible.   With    similar,  but   more  refined     criteria  for
distinguishing ``warm'' and ``cool'' galaxies, {}\scite{mann:warmcool}
find only a small  dependence of clustering on  the temperature in the
QDOT survey.
\item
We excluded  local  structures by looking  at  galaxies which  have  a
distance from the  galactic plane exceeding  20\hMpc.  We also changed
the    window  accordingly. The  distinct   features  remain,  but the
significance decreases.
\end{itemize}
Therefore,  the    morphometric difference between  the   northern and
southern  parts may not  be attributed  to  luminosity effects, colour
effects, or biases due to some peculiar local features.
\begin{itemize}
\item
By using volume limited subsamples we exclude  quite a few galaxies in
order   to get  ``clean''   samples  for   a   reliable analysis  with
geometrical  estimators. To check this dilution  effect we generated a
sequence of volume limited samples with limiting depths of 40\hMpc, 60
\hMpc, 80 \hMpc, 100 \hMpc, 150 \hMpc, 200\hMpc, and 300\hMpc.
\end{itemize}
Apart from the  300\hMpc\ sample, which  is too strongly diluted to be
significant, all  these  samples show significant  differences between
the northern and southern parts.
\begin{itemize}
\item
By   cutting  the  volume  limited   sample    with  100\hMpc\   depth
perpendicular to the galactic plane, we obtain four smaller parts with
approximately  160 galaxies   each. All  four  samples show  different
clustering properties. Therefore, the  detected fluctuations cannot be
attributed to a special north--south anisotropy.
\end{itemize}
In {}\scite{kerscher:significance} we  show   plots of  the  Minkowski
functionals  of  the   samples incorporating the   discussed selection
effects.

\subsubsection{Comparison with a simulation}
\label{sec:simulation}

\begin{figure}
\begin{center} 
\epsfxsize=7cm
\begin{minipage}{\epsfxsize}\epsffile{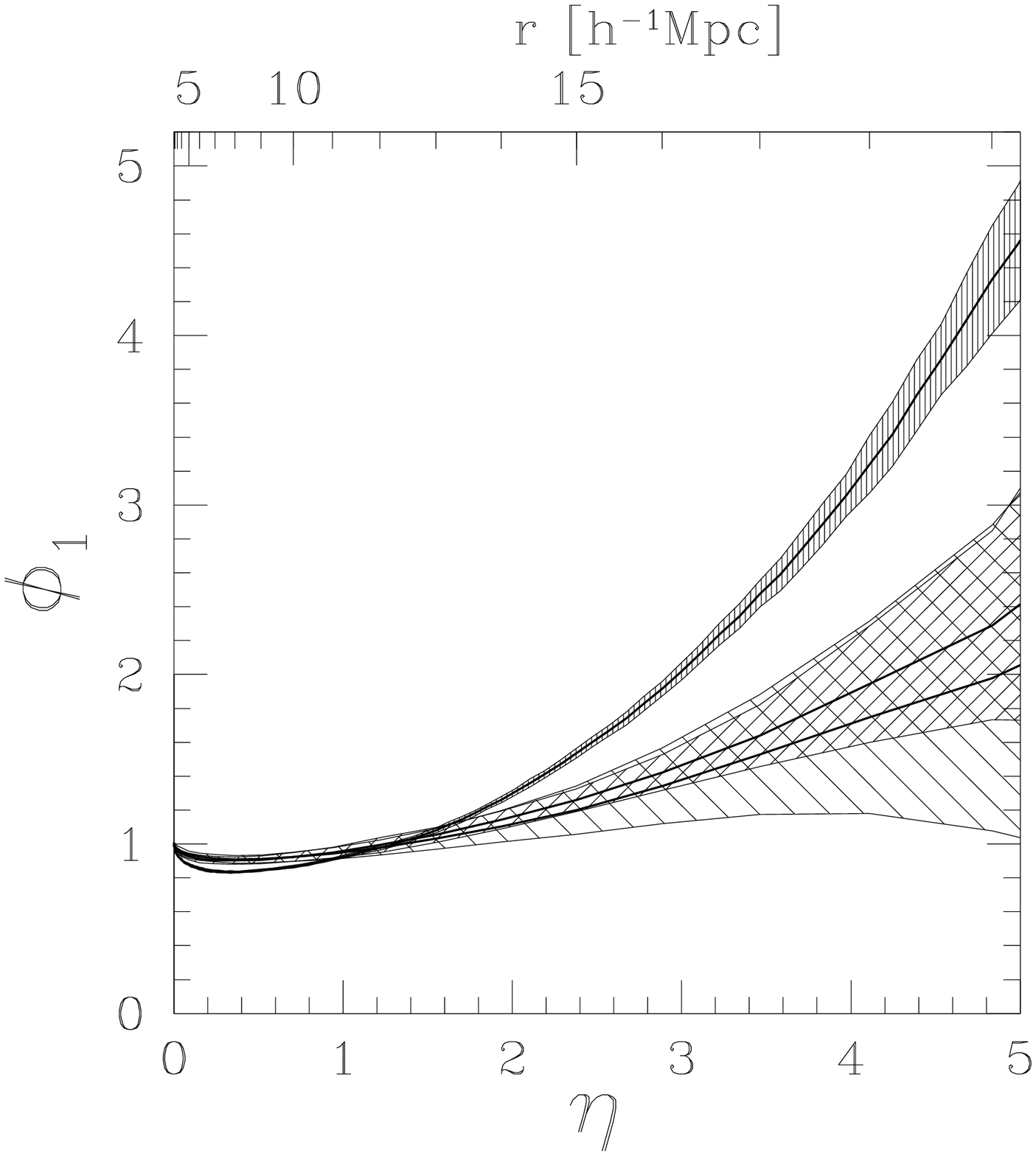}\end{minipage}
\epsfxsize=7cm
\begin{minipage}{\epsfxsize}\epsffile{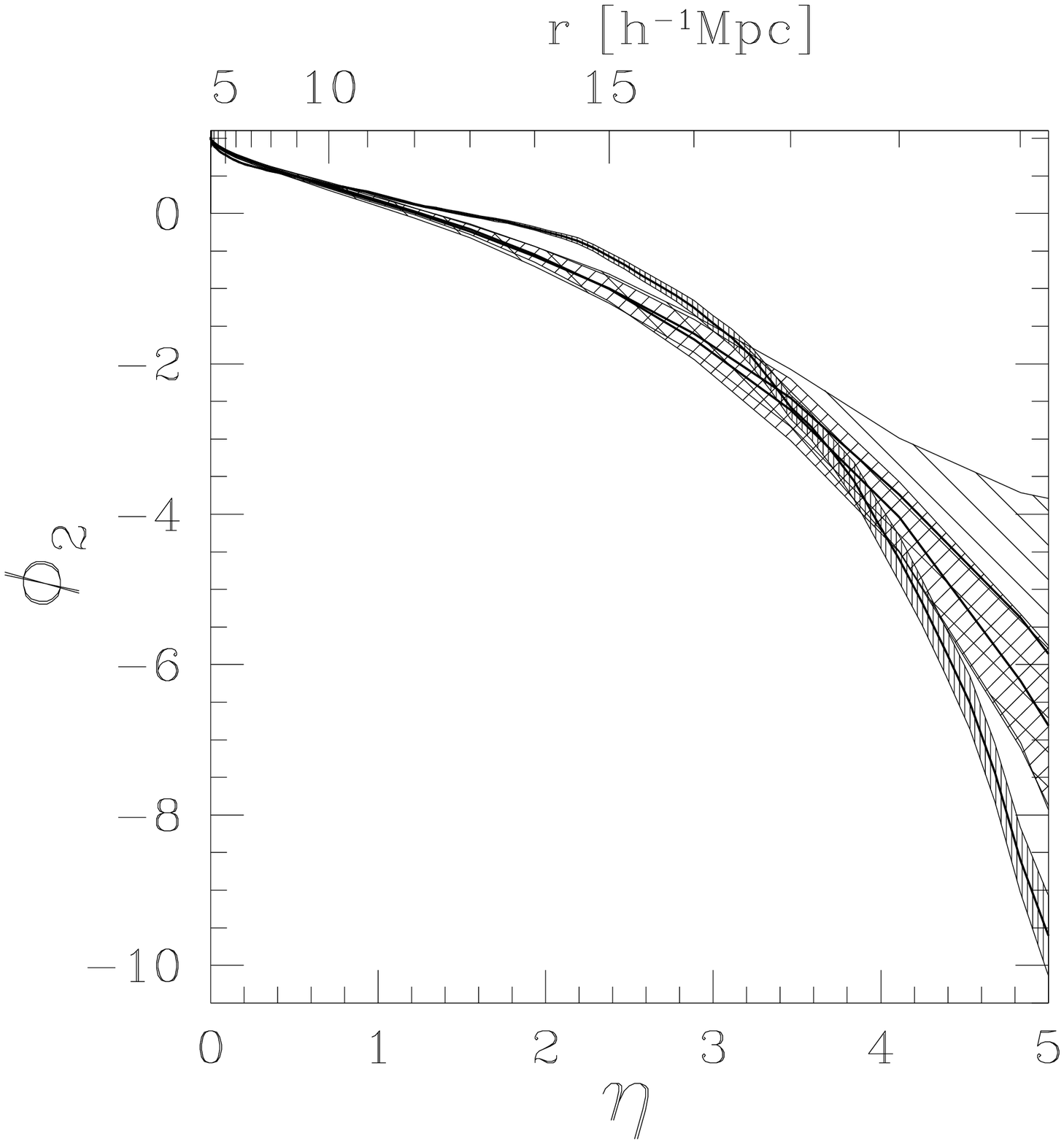}\end{minipage}
\epsfxsize=7cm
\begin{minipage}{\epsfxsize}\epsffile{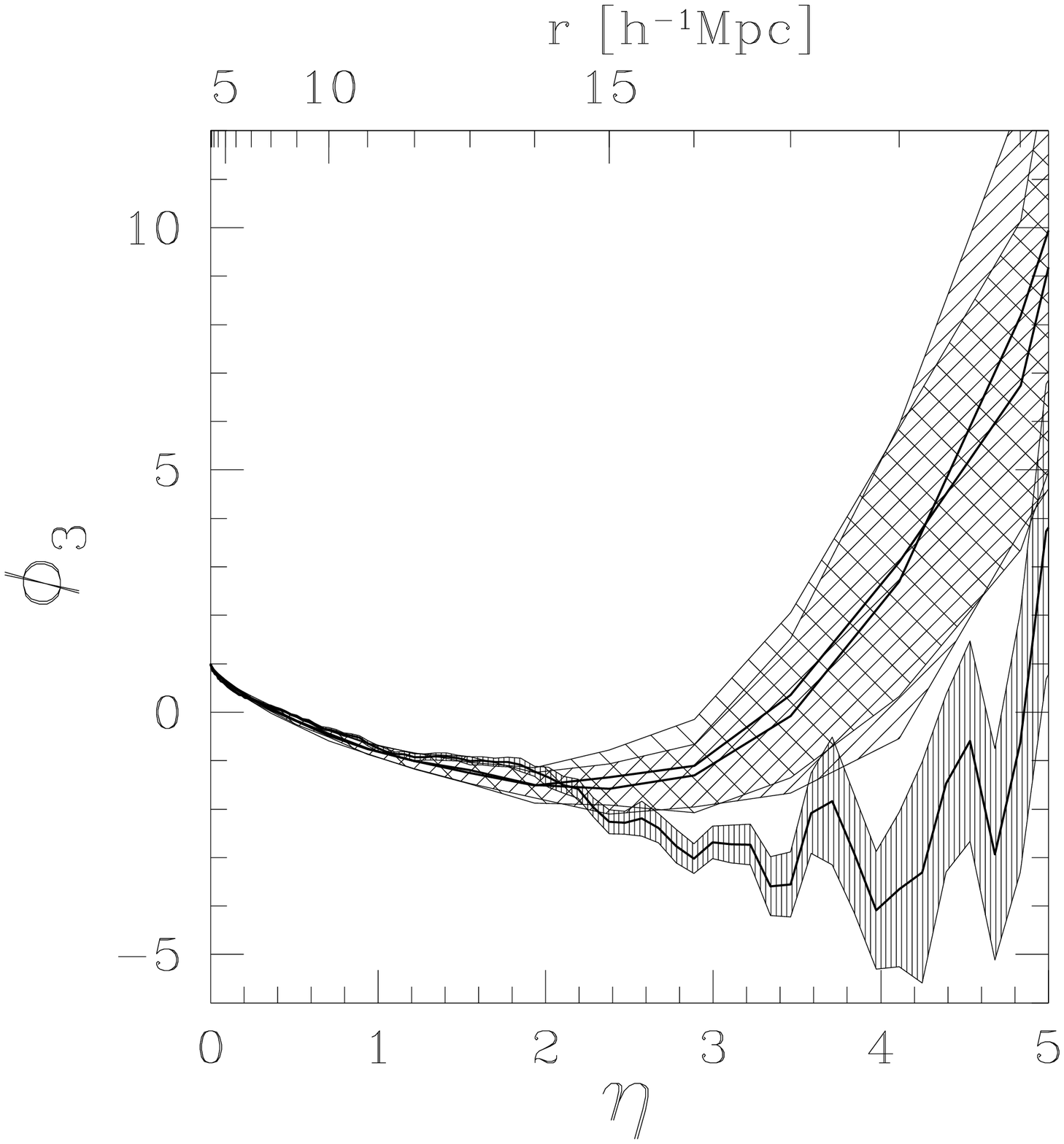}\end{minipage}
\end{center}
\caption{\label{fig:simulation} Minkowski  functionals $\phi_\mu$ of a
volume   limited sample with  $100\hMpc$  depth  and of mock  samples,
southern parts only.  The light shaded areas  represent the results in
redshift--space, the medium shaded the results in real--space, and the
dark shaded the results for the IRAS data.}
\end{figure}

To generate  mock  catalogues we use  the   results of  a cosmological
simulation  {}\cite{kolatt:simulating} kindly  provided  by Yair Sigad
and  Avishai  Dekel. The  simulation was  performed  for an $\Omega=1$
universe with initial conditions given by a constrained realization of
a Gaussian  random   field. The constraints   were  calculated from  a
density field  constructed from  the  IRAS 1.2~Jy galaxy catalogue  at
present  time,    evolved backwards  in  time   by  inversion  of  the
`Zel'dovich approximation'.   The   simulation  forward  in   time was
performed using   a particle--mesh code   with $128^3$ particles, each
comparable in mass to  an average galaxy,  in  a box with side  length
256\hMpc.   The purpose of this   simulation is to  mimick the current
distribution of  matter and  to  serve as  a testcase  for statistical
methods.  We constructed a sequence of  twenty mock catalogues both in
real--    and redshift--space from   this  simulation  by taking  only
galaxies which are inside  the  domain of  the sample  with  100\hMpc\
depth  (see  Fig.~{}\ref{fig:window}).   By randomly  selecting  these
galaxies we make sure to  have the same  number density as observed in
the  data.  Since we want to  compare  the morphological properties of
the mock samples with the observed galaxy distribution in the northern
and southern  part separately, we choose  the same orientation  of the
galactic plane as used for the  constrained realization of the initial
density field.

In  Fig.~\ref{fig:simulation}  we  show  the  average   values  of the
Minkowski functionals  for the southern  part of the mock samples both
in real-- and   redshift--space   in comparison   with the   Minkowski
functionals of the volume   limited  observed sample with  100  \hMpc\
depth. For the simulation the error  is found by averaging over twenty
mock  catalogues extracted from the same  simulation, for  the data we
plot   the subsampling   error.     Both  results,  from  real--   and
redshift--space,  are     consistent    within   their  errors    (the
redshift--space   results  show larger scattering),  but  they clearly
differ from the IRAS data.  We attribute this to the small size of the
simulation  box (256\hMpc\ sidelength) compared  to the  extent of the
mock samples (200\hMpc\ diameter). The simulation enforces homogeneity
on the scale of the sidelength of  the box, suppresses fluctuations on
that scale, and therefore is not  able to reproduce varying clustering
properties in distinct regions which nearly fill half of the box each.

Obviously, we demand too  much from the  simulation which was designed
to reproduce the  density  and velocity  fields (and especially  their
relation) in the  nearby  region of the   universe out to  60\hMpc. In
order to cover fluctuations  on the scale  of 200\hMpc, the simulation
simply needs to be significantly larger.

In the case of the northern part the results for the IRAS data and the
mock catalogues agree   (we do not  show  them here).  The  real-- and
redshift--space results now show the same scattering.  The consistency
of real-- and redshift--space results,  both in north and south, gives
us confidence that our analysis of the  observed IRAS 1.2~Jy catalogue
is not affected by redshift space distortions.

\subsubsection{Nearest neighbour statistics and $\sigma^2$}
\label{sec:nn_sigma_v10}

\begin{figure}
\begin{center} 
\epsfxsize=7cm
\begin{minipage}{\epsfxsize}
\epsffile{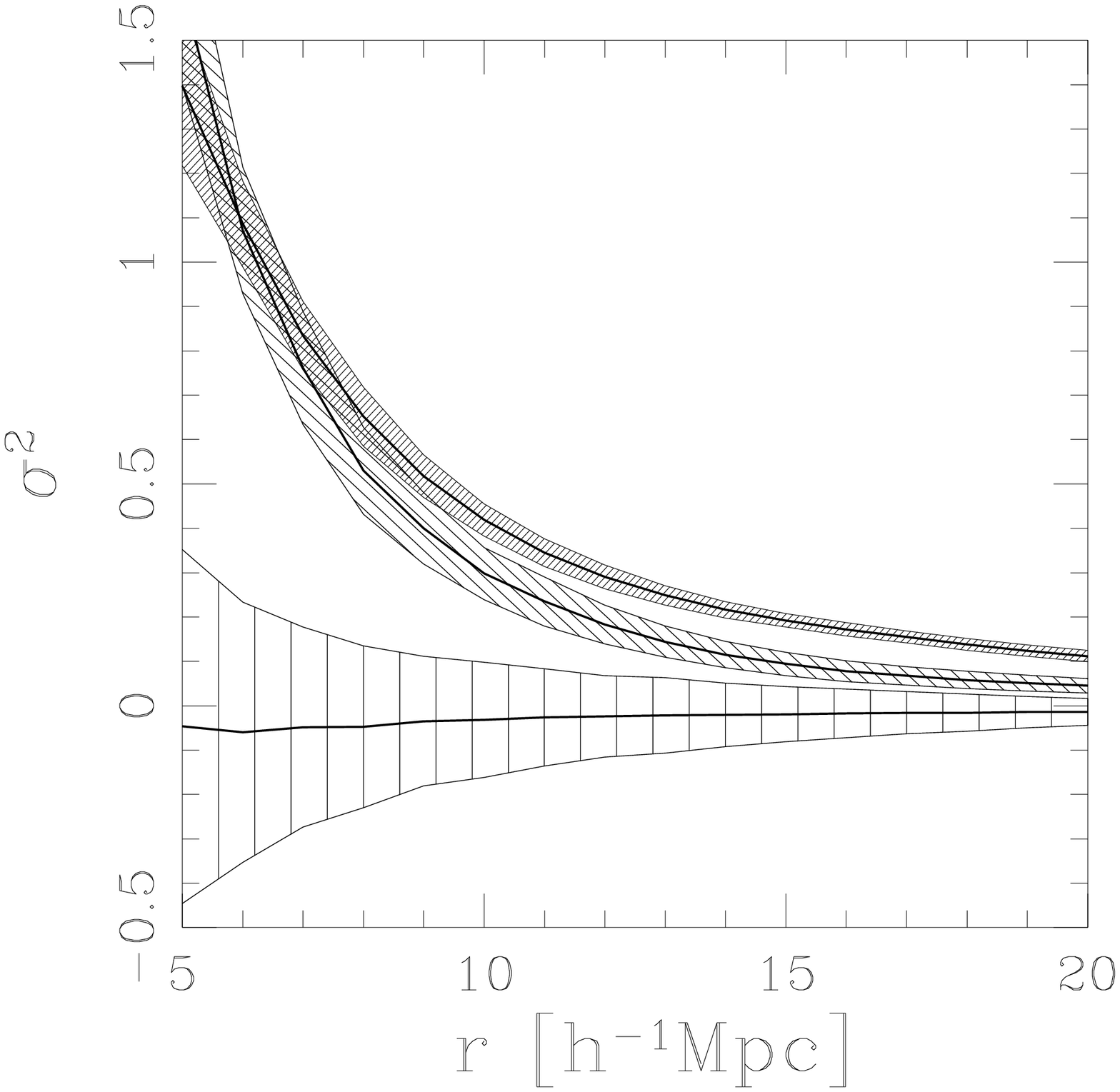}
\end{minipage}
\end{center}
\caption{\label{fig:sigma_v10} $\sigma^2$    for  the  volume  limited
samples with 100\hMpc\  depth; again, the  dark shaded areas represent
the southern part, the medium shaded the northern  part, and the light
shaded a Poisson process with the same number density.}
\end{figure}

\begin{figure}
\begin{center} 
\epsfxsize=7cm
\begin{minipage}{\epsfxsize}\epsffile{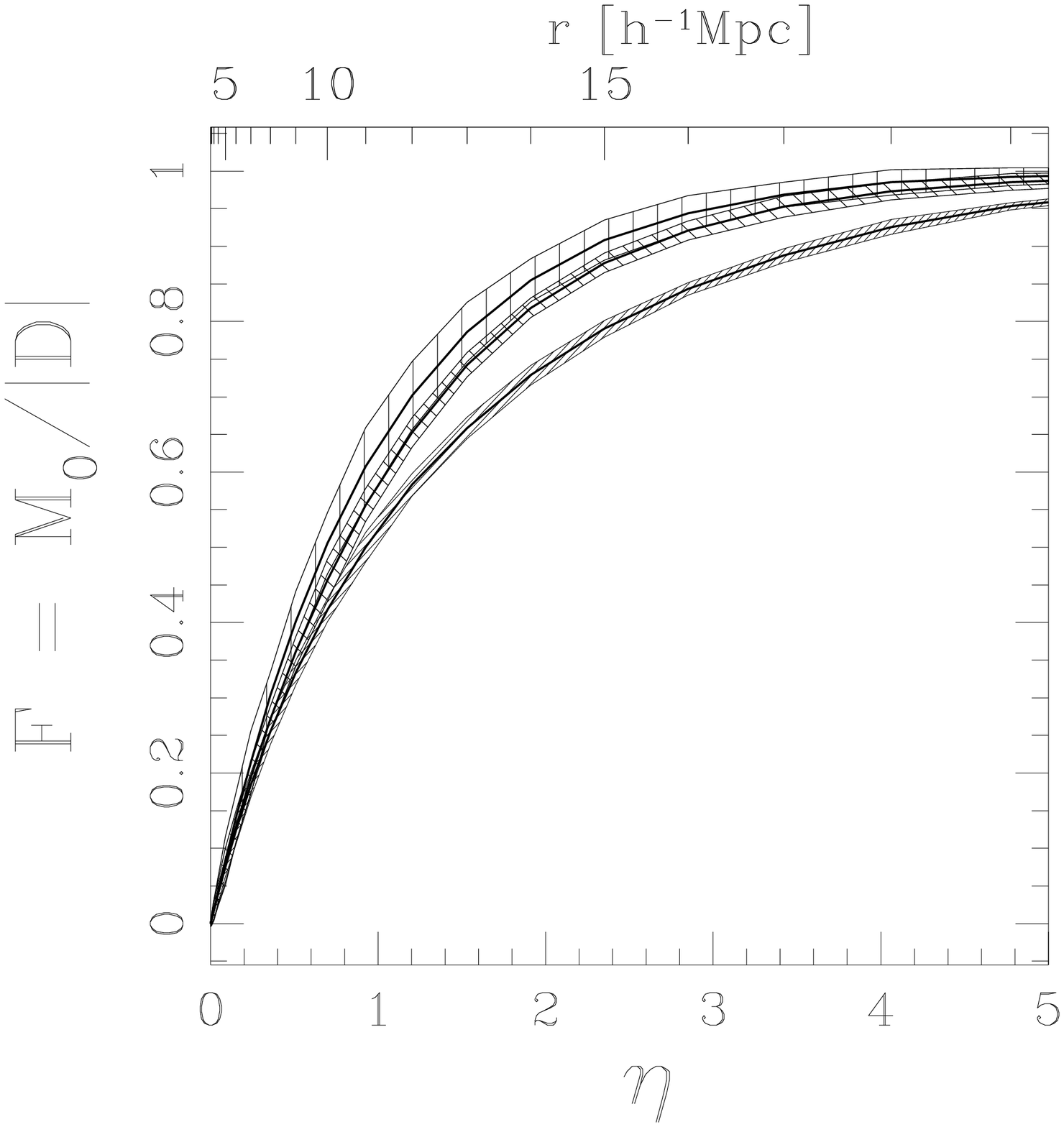}\end{minipage}
\epsfxsize=7cm
\begin{minipage}{\epsfxsize}\epsffile{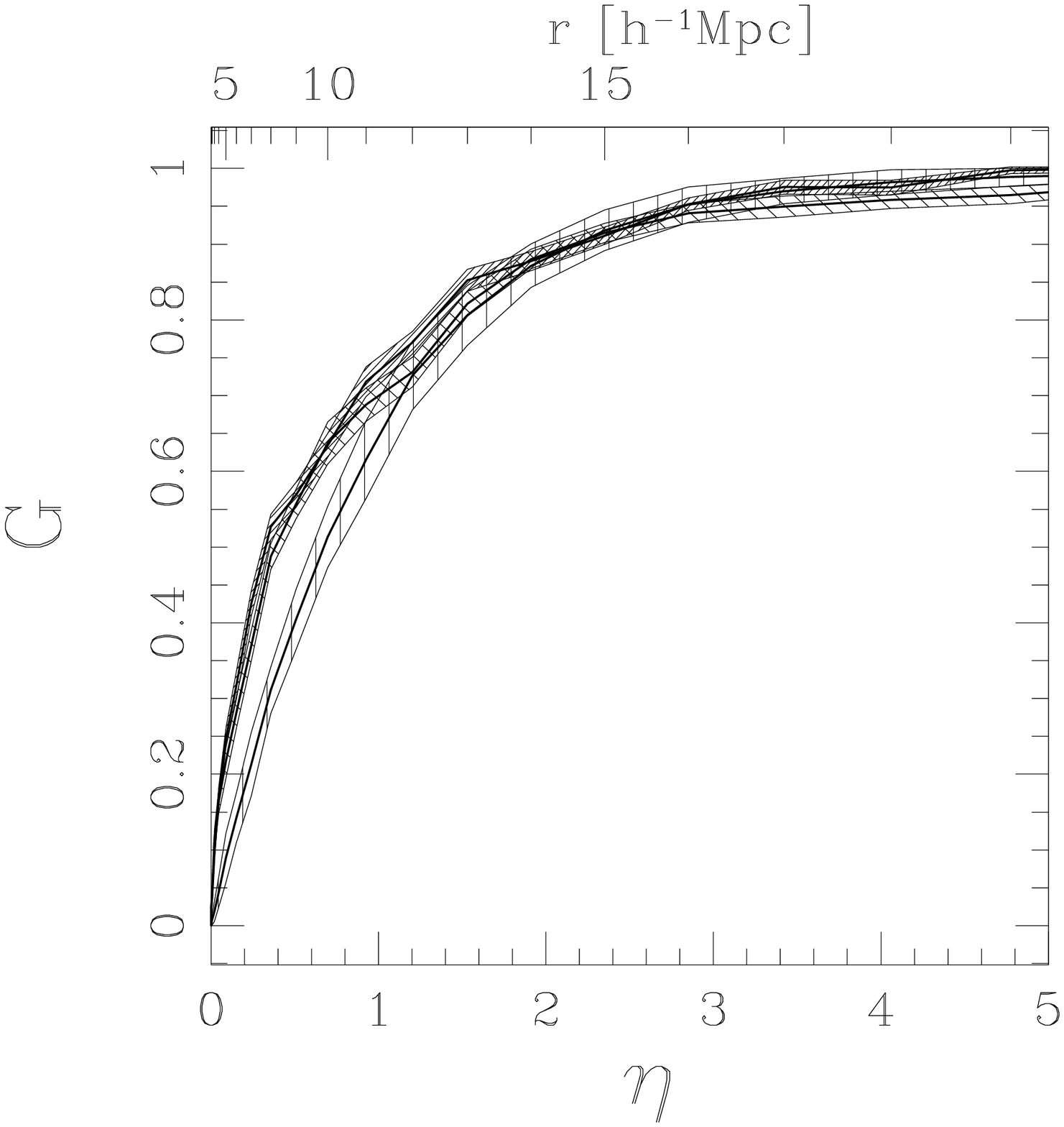}\end{minipage}
\epsfxsize=7cm
\begin{minipage}{\epsfxsize}\epsffile{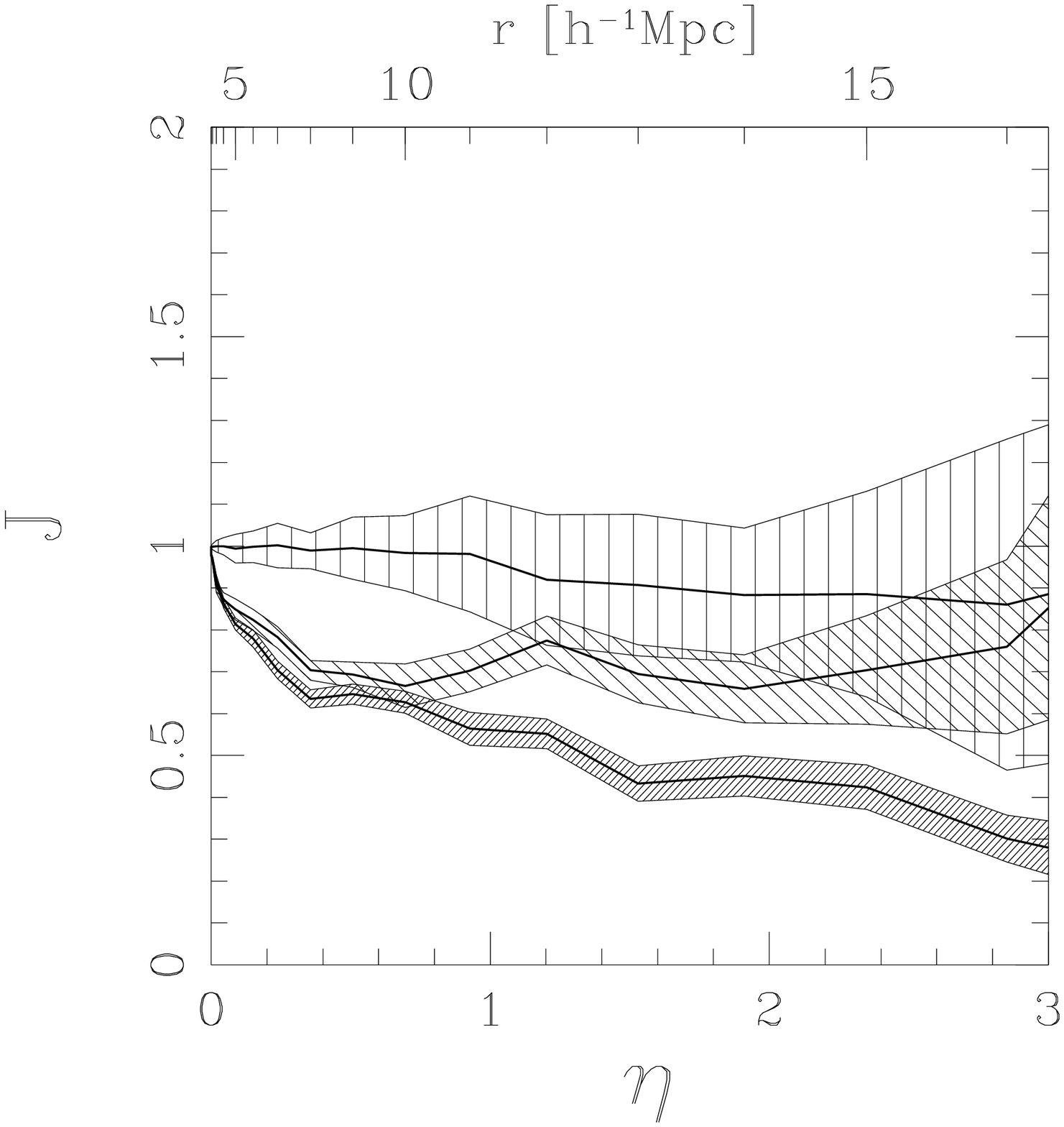}\end{minipage}
\end{center}
\caption{\label{fig:FGJjyv10} $F$, $G$  and $J$ for the volume limited
samples  with 100\hMpc\  depth;  the dark  shaded areas  represent the
southern part,  the  medium shaded the northern   part, and the  light
shaded a Poisson process with the same number density.}
\end{figure}

In Fig.~(\ref{fig:FGJjyv10}) we  show the empty space function $F(r)$,
the nearest neighbour  distribution  $G(r)$ and the $J(r)$  statistics
for the volume limited sample with  100\hMpc. The errors are estimated
in the same way as for the  Minkowski functionals. The error increases
with the   radius,  since $F$   and  $G$ are approaching   unity,  and
$(1-G)/(1-F)$ diverges due to random fluctuations.

The   volume densities $F(r)$ differ  significantly  between north and
south. The lowered volume density in the  southern part on scales from
10 to 20\hMpc\  indicates stronger clustering  than in  the case  of a
Poisson process, while the northern part is marginally consistent with
a  Poisson process.    The  nearest  neighbour distributions  of   the
northern and  southern  parts practically overlap,  but differ clearly
from the  Poisson process.  On scales  from  10  to 15\hMpc\  the  $J$
statistics  discriminates  between  northern part, southern   part and
Poisson process, again  showing stronger  clustering (reduced $J$)  in
the south.

In Fig.~(\ref{fig:sigma_v10}) we   present the results for  the galaxy
count fluctuations in a  volume limited sample with\  100\hMpc\ depth.
On the scale of 8\hMpc\ both parts  display comparable fluctuations of
counts in cells with $\sigma_8:=\sigma(8\hMpc)=0.81\pm0.06$ ($1\sigma$
error,    southern      part)   and $\sigma_8=0.73\pm0.07$   (northern
part). These     directly   measured  values  both    agree   with the
$\sigma_8=0.77\pm0.19$   determined by {}\scite{fisher:clusteringI} in
redshift space,  using Eq.~(\ref{eq:sigma-powerlaw})  and a power--law
fit  to the correlation function.   On   larger scales the  difference
between north and south  is clearly seen, as  well as  their deviation
from a Poisson process.

\subsection{Volume limited samples with 200\hMpc\ depth}
To look deeper into space we use  a sample which  is volume limited to
200\hMpc\ depth. The northern part then contains 139 galaxies, and the
southern part 141  galaxies. The redshift   and flux distribution  are
similar to those shown in Fig.~\ref{fig:jyv10-hist}.

\subsubsection{Minkowski functionals}
\begin{figure}
 \begin{center}
 \epsfxsize=7cm
 \begin{minipage}{\epsfxsize} \epsffile{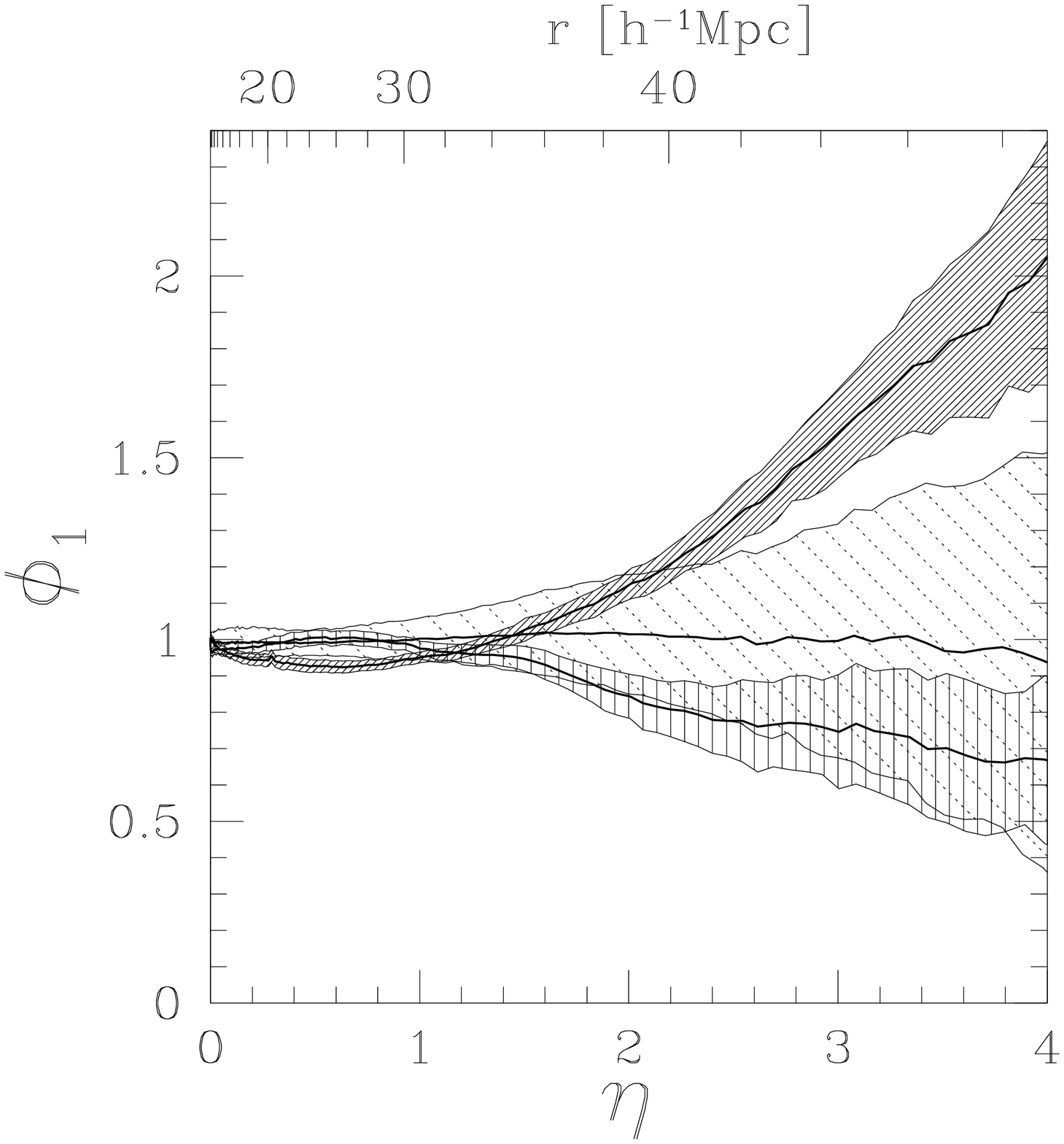} \end{minipage} 
 \epsfxsize=7cm
 \begin{minipage}{\epsfxsize} \epsffile{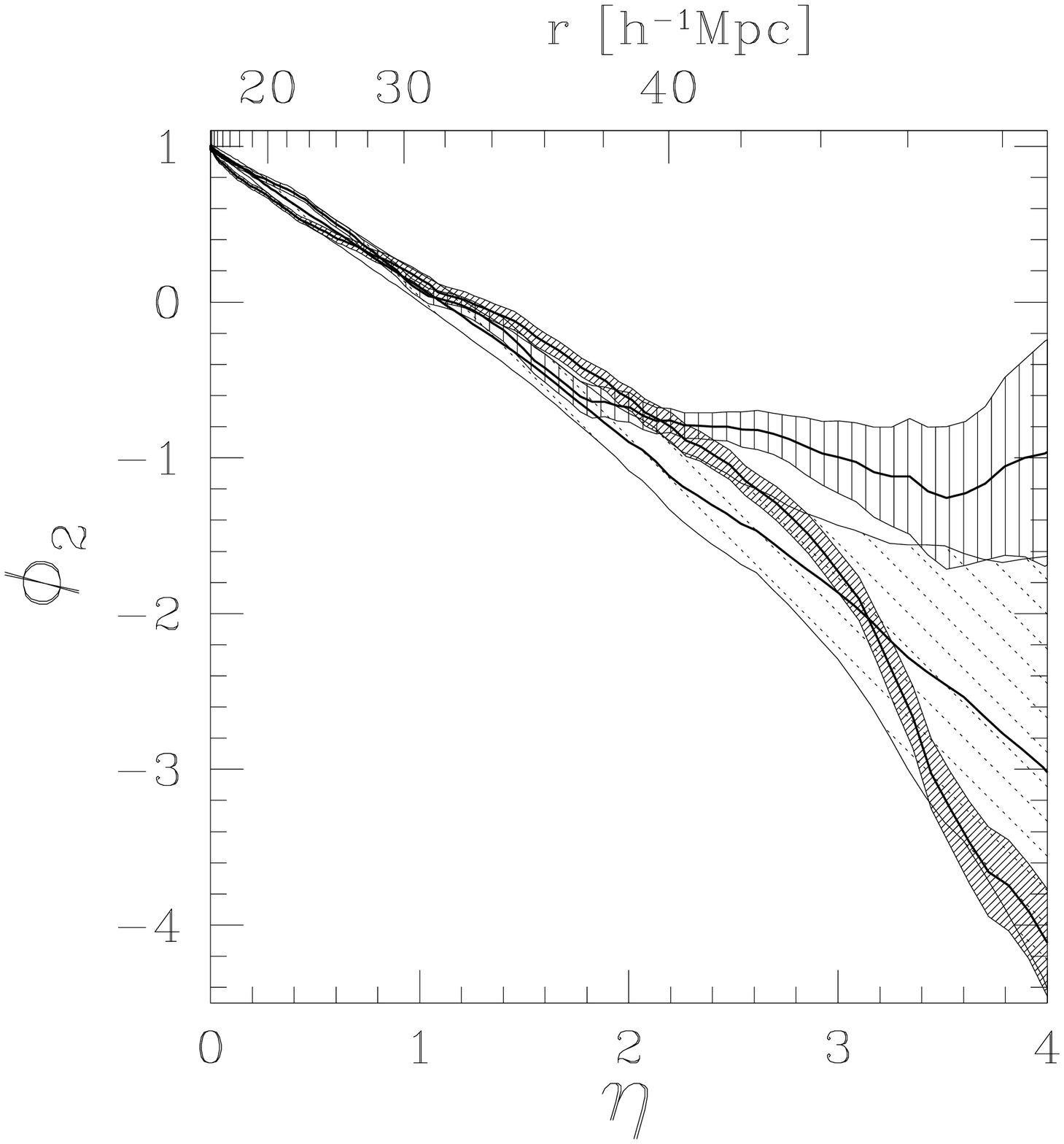} \end{minipage} 
 \epsfxsize=7cm
 \begin{minipage}{\epsfxsize} \epsffile{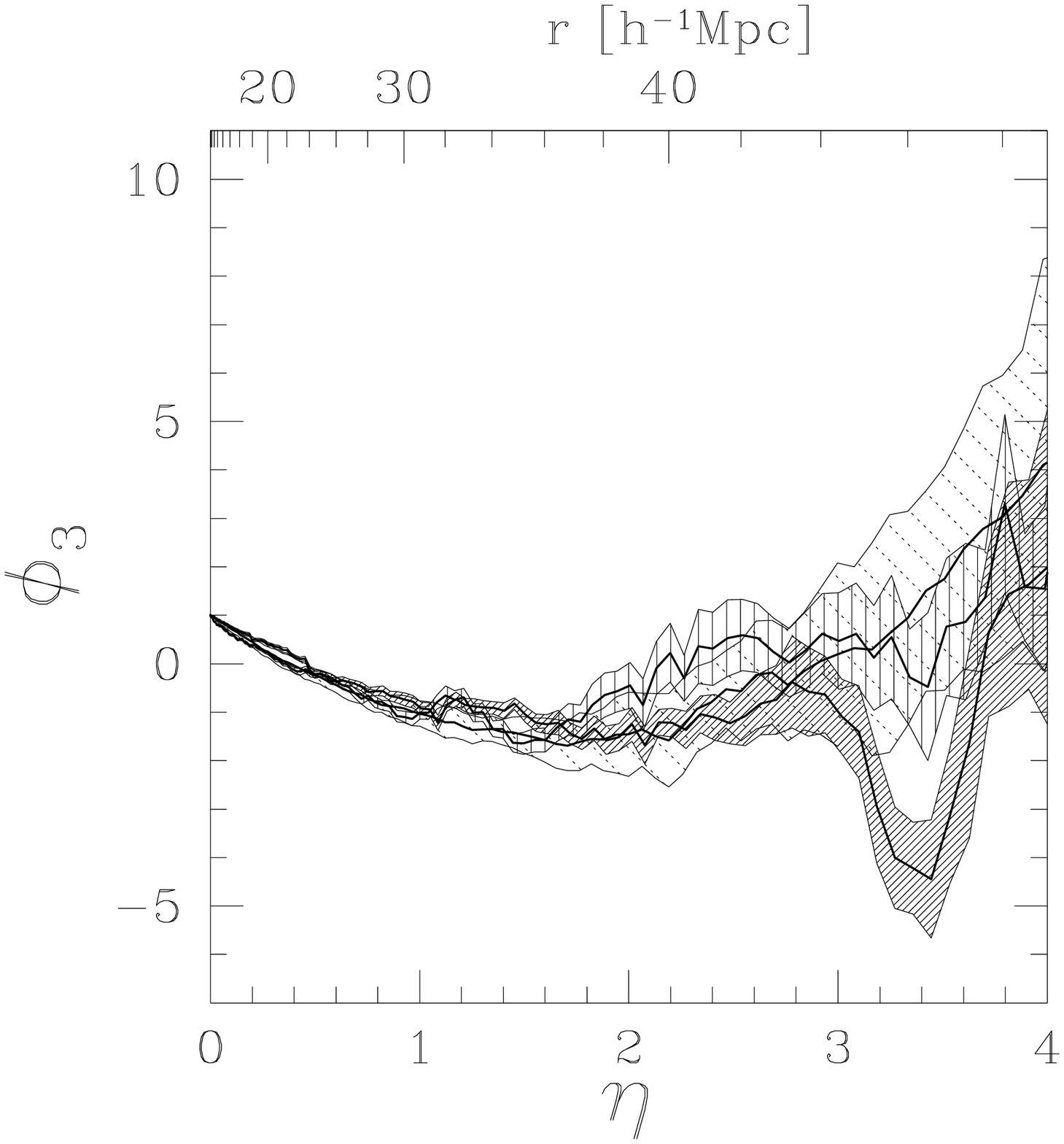} \end{minipage} 
 \end{center}
\caption{\label{fig:minjyv20}           Minkowski          functionals
$\phi_\mu(\cA_N(r))$ of a volume limited  sample with 200\hMpc\ depth;
the dark  shaded areas represent  the southern part, the medium shaded
the northern  part, and the  light shaded  a Poisson  process with the
same number density.}
\end{figure}

In Fig.~\ref{fig:minjyv20} the  values of the Minkowski functionals of
the   southern and  northern  parts   are  plotted together with   the
functionals of  a  Poisson process with  the  same  number density. As
before, the errors are determined via subsampling with 90\%.
Again, the southern part shows a tendency towards stronger clustering,
and the  enhanced  surface area is still   clearly detectable, but the
overall   uncertainty has increased  considerably.  As  pointed out in
Sect.~\ref{sec:selection} the reduced deviations from the Poisson data
should not be attributed to a homogenization  of the underlying matter
distribution on the scale  of 200\hMpc, but rather  are due to  sparse
sampling with only $\approx 140$ galaxies in each part.

\subsubsection{Nearest neighbour statistics and $\sigma^2$}
\label{sec:nn_sigma_v20}
\begin{figure}
 \begin{center} 
 \epsfxsize=7cm
 \begin{minipage}{\epsfxsize}\epsffile{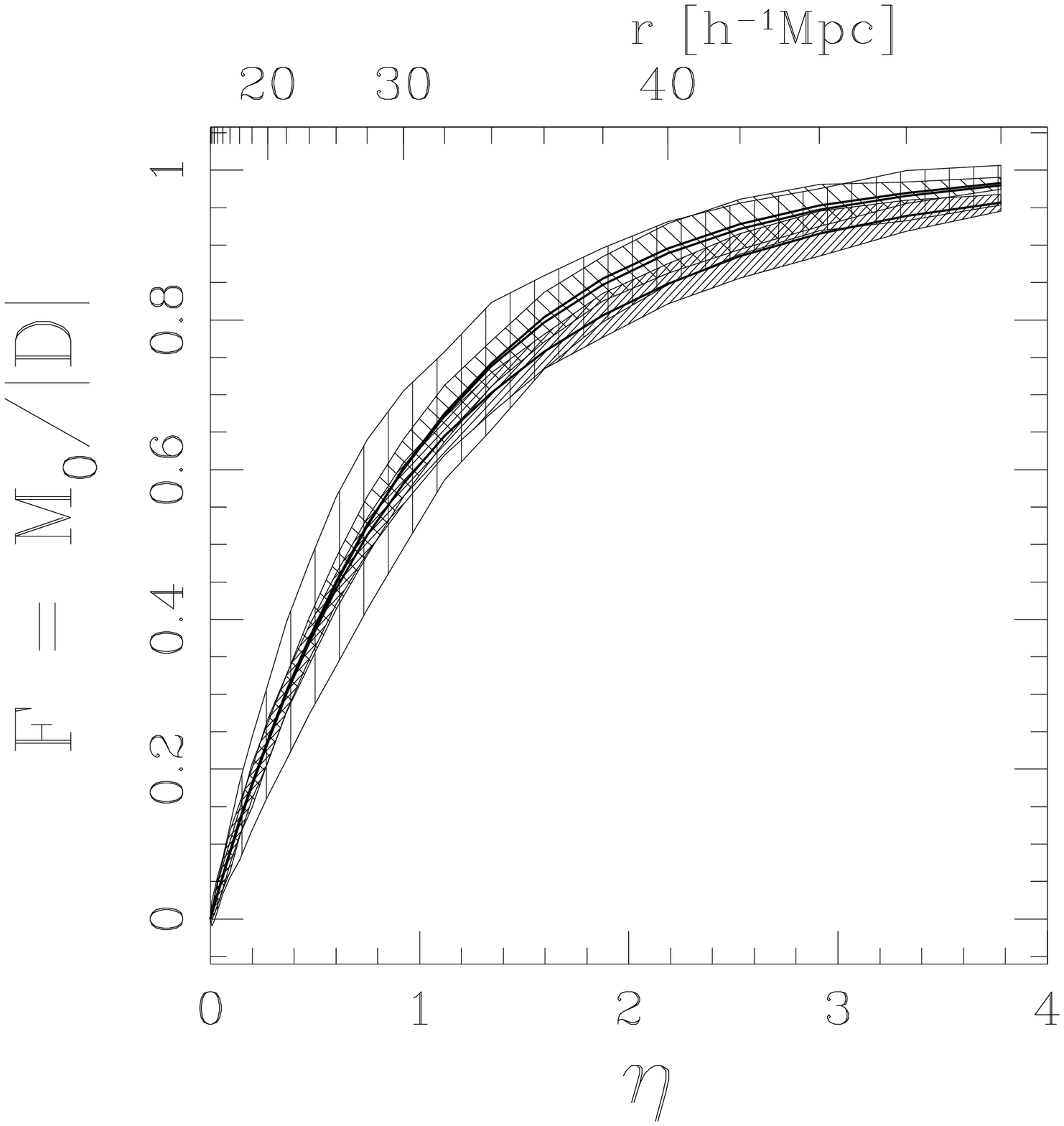}\end{minipage}
 \epsfxsize=7cm
 \begin{minipage}{\epsfxsize}\epsffile{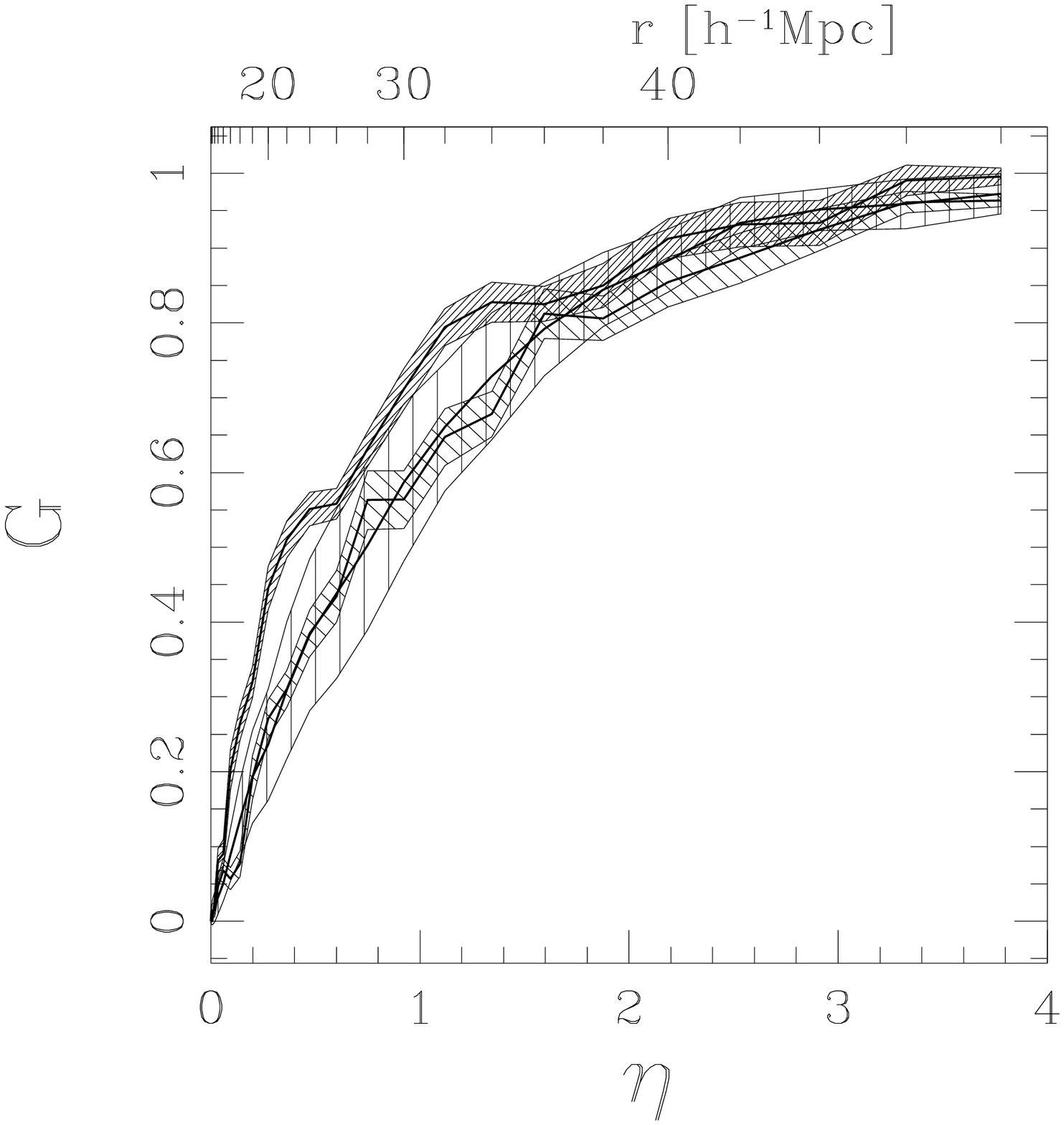}\end{minipage}
 \epsfxsize=7cm
 \begin{minipage}{\epsfxsize}\epsffile{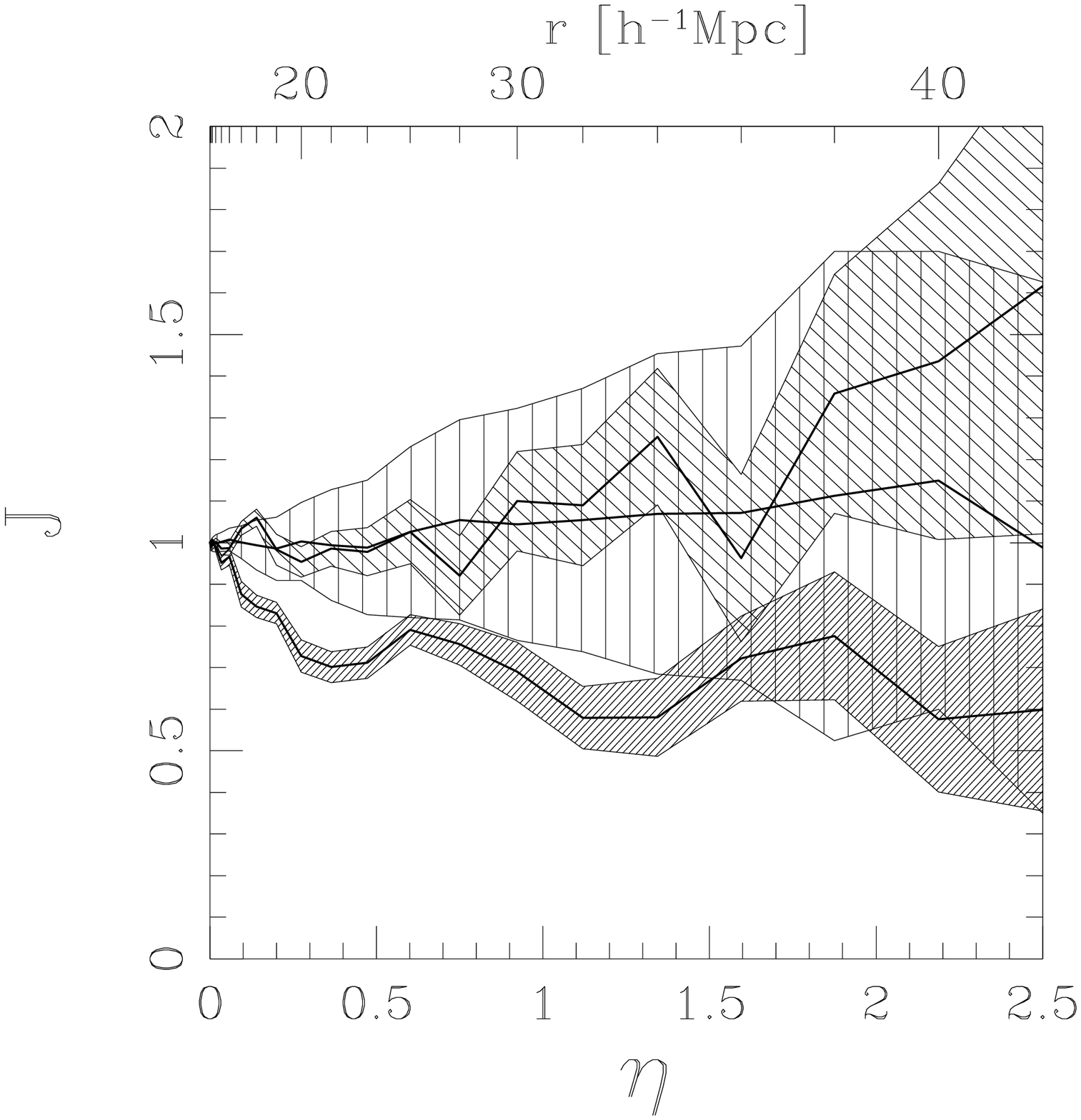}\end{minipage}
 \end{center}
\caption{\label{fig:GJjyv20} $F$,  $G$  and $J$  of  a volume  limited
sample   with  200\hMpc\ depth; the  dark   shaded areas represent the
southern  part, the  medium shaded  the  northern part, and the  light
shaded a Poisson process with the same number density.}
\end{figure}

The nearest  neighbour distribution $G(r)$   and the $J(r)$ statistics
for  the volume limited   sample with  200\hMpc\  depth are  shown  in
Fig.~\ref{fig:GJjyv20}.
The volume density, the nearest neighbour distribution, and the $J(r)$
statistics of  the  northern   part are  consistent  with  a   Poisson
distribution, while  the southern part  differs  in $G(r)$  and $J(r)$
both from a Poisson process and from the northern part.

\begin{figure}
\begin{center} 
\epsfxsize=7cm
\begin{minipage}{\epsfxsize}\epsffile{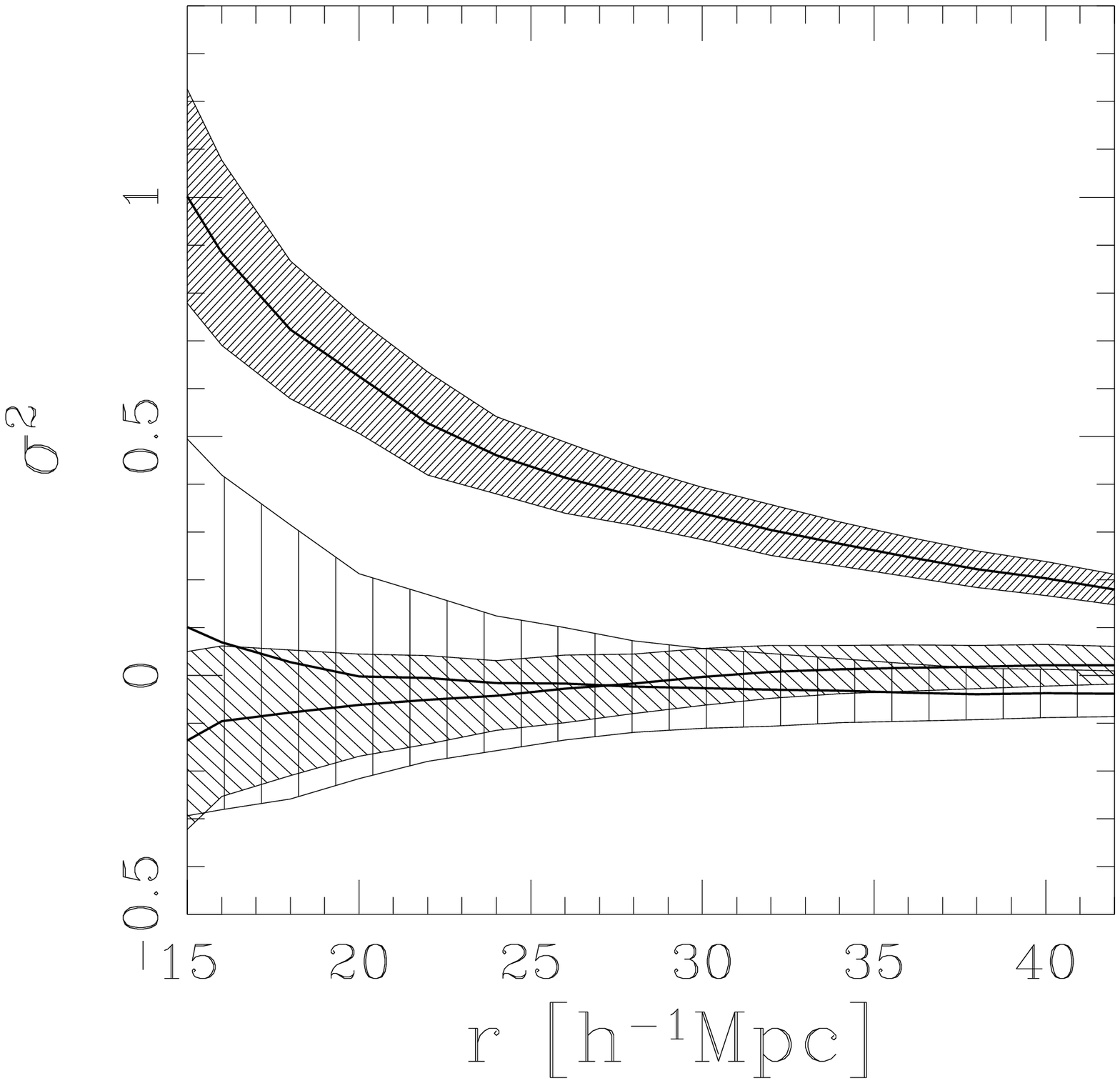}\end{minipage}
\end{center}
\caption{\label{fig:sigma_v20}   $\sigma^2$   for the   volume limited
samples with 200\hMpc\ depth;  again, the dark shaded areas  represent
the southern part, the medium shaded the northern  part, and the light
shaded a Poisson process with the same number density.}
\end{figure}

The galaxy count fluctuations  (Fig.~\ref{fig:sigma_v20}) for a volume
limited sample with 200\hMpc\ depth show  clearly a difference between
northern and southern parts, but only the  southern part deviates from
a Poisson process.

\begin{figure}
\begin{center} 
\epsfxsize=7cm
\begin{minipage}{\epsfxsize}\epsffile{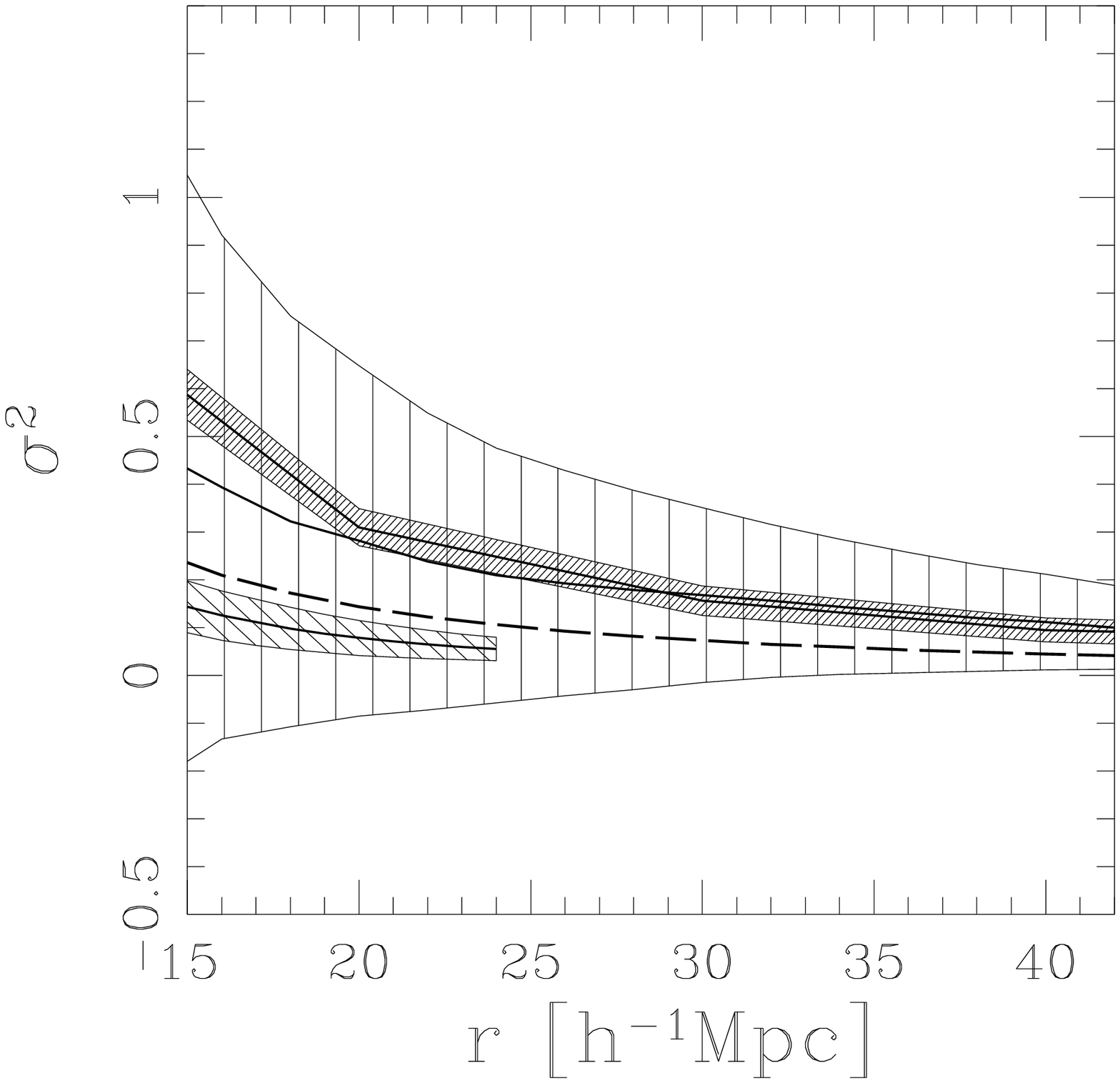}\end{minipage}
\end{center}
\caption{\label{fig:sigma_comb} The mean values for $\sigma^2$ for the
volume limited samples with   100\hMpc\ and 200\hMpc\ depth;  the dark
shaded  area marks  the result  of {}\protect\scite{oliver:largeiras},
the medium shaded the  mean result for  the 100\hMpc\ sample,  and the
light shaded the mean result for the 200\hMpc\ sample. The dashed line
gives the results  for $\sigma^2$ as in  Eq.~(\ref{eq:sigma-powerlaw})
with    $r_0=4.27\hMpc$    and       $\gamma=1.68$     according    to
{}\protect\scite{fisher:clusteringI}.}
\end{figure}

Figure~{}\ref{fig:sigma_comb}   presents   the    average   values  of
$\sigma^2$  over   both parts  for  volume  limits   of 100\hMpc\  and
200\hMpc\ together with  the results of {}\scite{oliver:largeiras} and
{}\scite{fisher:clusteringI}.
For  the 200\hMpc\  sample our averaged   values are in good agreement
with   the  results of  {}\scite{oliver:largeiras};  since  $\sigma^2$
differ between the  northern and southern parts,  the errors quoted by
{}\scite{oliver:largeiras} are underestimated. This may be a result of
their method:  These authors calculate $\sigma^2$   in shells, and use
the mean number density {\em in one shell} in calculating the variance
of the  count  in  cells. Hence,   they  {\em assume}   homogeneity by
implicitly weighting with the selection function.

We also plot  $\sigma^2(r)$ according to Eq.~(\ref{eq:sigma-powerlaw})
with the   values  $r_0 =  4.27^{+0.66}_{-0.81} \hMpc$  and  $\gamma =
1.68^{+0.36}_{-0.29}$ as given by {}\scite{fisher:clusteringI} for the
volume  limited  sample with  100\hMpc\  depth  for both  northern and
southern parts.  These values  were  determined  from   a fit to   the
correlation function in the range from 1 to 13\hMpc. Within the quoted
errors   our   results     are  consistent   with    the   results  of
{}\scite{fisher:clusteringI},   with  a clear  tendency  towards lower
values  of $\sigma^2$   on larger scales  (compared   to the 100\hMpc\
sample).

The  $\sigma^2$   determined from the    volume  limited samples  with
$100\hMpc$ and $200\hMpc$  differ.   This variation of $\sigma^2$  for
the 1.2~Jy data reveals a dependency  of $\sigma^2$ on the mean number
density  $\overline{n}$    and    on   the  depth   of     the samples
($\overline{n}=1.84\times10^{-4}\, (\hMpc)^{-3}$  for the sample  with
100\hMpc\ depth and $\overline{n}=9.05\times10^{-6}\,(\hMpc)^{-3}$ for
the sample  with 200\hMpc\ depth).  In  Appendix~B we show for a point
distribution on  a fractal, as   a simple still  tractable model, that
$\sigma^2$ depends  on the mean number  density $\overline{n}$ and the
depth of the sample.

\section{Morphometry of a point set}
In   previous    sections,  we   applied    Minkowski  functionals  as
morphological tools to analyze   spatial patterns displayed by  galaxy
catalogue data. Here,   we comment on  the  rationale of this  method,
which became established only recently in the cosmology literature.

Basically,  a galaxy   catalogue  is a finite  set  of triple--numbers
$X=\{\bx_i\}_{i=1}^N$  representing  a  skeleton   pattern   of points
imbedded in a domain $D$ of a  three--dimensional Euclidean space. The
geometrical information supplied by  this skeleton may be organized in
a collection of particle distributions,
\begin{equation}
\Gamma_n(\by_1,\dots,\by_n|D)~,\quad1{\le}n{\le}N,
\end{equation}
with
\begin{multline}
\Gamma_n(\by_1,\dots,\by_n|D):= \\
\sum_{(i_1,\ldots,i_n)} 
\delta(\by_1-\bx_{i_1})\dots\delta(\by_n-\bx_{i_n}),
\end{multline}
where the sum runs over all $n$--tupels $(i_1,\ldots,i_n)$ with
$1{\le}i_1{<}\ldots{<}i_n{\le}N$, and each $\by_i$ is chosen to lie within
$D$.

What is  the ``form'' of this point  set? Visual inspection habitually
conveys the impression of distinct structural features such as clumps,
voids, ``Great Walls'',  ``chessboard universe'', or even ``Fingers of
God''.  However,  these subjective  impressions  may be deceptive.  To
associate  a  form with  a   point set,  we   must  put flesh  on  the
skeleton.  We have done  this  in a   technically controllable way  by
covering the points with  copies of a ball $\cB_r$   to create a  body
$\cA_N(r)  =\bigcup_{i=0}^N\cB_r(\bx_i)$, where the  common radius $r$
may be employed as a variable diagnostic parameter.

The  Minkowski  functionals  are    directly  related with    familiar
geometrical   and         topological    quantities    (see      Table
{}\ref{table:mingeom}) and  are   ideally suited  to  measure content,
shape,  and  connectivity of  bodies such  as $\cA_N(r)$.  In order to
explicate their connection with the catalogue data, let us write
\begin{equation}
\cA_N(r)=\bigcup_{i=0}^N\cB_r(\bx_i)=\cA_{N-1}(r)\cup\cB_r(\bx_N).
\end{equation}
One of the basic properties of the  functionals $M_\mu(\cdot)$ is {\em
additivity} which implies, in particular,
\begin{multline}\label{eq:min_additivity}
M_\mu(\cA_N(r))=M_\mu(\cB_r(\bx_N))+M_\mu(\cA_{N-1}(r))-\\
M_\mu(\cA_{N-1}(r)\cap\cB_r(\bx_N)).
\end{multline}
By iterating Eq.~(\ref{eq:min_additivity}), we ultimately arrive at
\begin{multline}\label{eq:additivity_iterated}
M_\mu(\cA_N(r))=\\
\sum_{n=1}^N (-1)^{n-1} \sum_{(i_1,\dots,i_n)} 
M_\mu(\bx_{i_1},\dots,\bx_{i_n};r) ,
\end{multline}
where    we    have      set    $M_\mu(\bx_{i_1},\dots,\bx_{i_n};r)  =
M_\mu(\cB_r(\bx_{i_1})\cap\dots\cap\cB_r(\bx_{i_n}))$.   In  terms  of
the     particle    distribution     $\Gamma_n$,    the     expression
(\ref{eq:additivity_iterated}) can be written as
\begin{multline}
M_\mu(\cA_N(r)) =
\sum_{n=1}^N(-1)^{n-1}\int_D\!\d^3\!x_1\dots\int_D\!\d^3\!x_n\\
M_\mu(\bx_1,\dots,\bx_n;r)\Gamma_n(\bx_1,\dots,\bx_n|D).
\end{multline}
Evidently,  the values $M_\mu(\cA_N(r))$  of the Minkowski functionals
are determined uniquely by the input data encoded in the distributions
$\Gamma_n$,    once         a      covering       is           chosen.
Equation~(\ref{eq:min_additivity}) remains   valid when the  balls are
replaced   by congruent copies   of   some other compact convex   set;
however, the values $M_\mu(\cdot)$ for a given catalogue vary with the
choice of the coverage.

We    note  that the present  approach    is  purely combinatorial; no
statistical assumptions     enter (beyond  those  involved    in   the
construction of  the  sample catalogue),  and  the  problem  of ``fair
samples'' does not arise.

Let us compare two samples $(1)$ and $(2)$ with the number of galaxies
$N_1{\approx}N_2$  in  a  common domain.   If  we find the  values  of
$M_\mu(\cA^{(1)}_{N_1}(r))$  and $M_\mu(\cA^{(2)}_{N_2}(r))$ to differ
over some finite  range of values of the  radius $r$ then the  samples
are certainly distinct   morphologically. For instance, the  V--shaped
spike   in the  normalized  Euler   characteristic  $\Phi_3$, seen  in
Fig.~\ref{fig:minjyv20}, is  not a random deviation  due to  some poor
statistics  but provides definitive  evidence for  higher connectivity
(i.e.\ higher genus) of the southern  covering body in comparison with
the northern one when $r\approx46\hMpc$.

\section{Summary and Conclusions}

We analyzed  morphological characteristics of the  galaxy distribution
described by the IRAS 1.2~Jy catalogue.  The two subsamples (north and
south) of this catalogue  were studied individually by employing three
different methods: (i)  Minkowski functionals, (ii) nearest--neighbour
distribution, and (iii) variance in the galaxy number counts.

Since the IRAS  data have been obtained  from a single instrument with
uniform calibration, the two  subsamples, which contain about the same
number of galaxies, can be compared reliably.

To assess the significance of our results we took some care to account
for selection and finite size effects.  For reference purposes we used
typical realizations of a stationary Poisson point process.

Our results may be summarized as follows:
\begin{itemize}
\item
The values of the  Minkowski functionals for  the southern part differ
significantly from those  for the  northern  part. This  morphological
segregation is seen  in    the volume  limited  subsamples both   with
100\hMpc\ depth and with 200\hMpc\ depth, but it is less pronounced in
the latter  case because of sparse  sampling.  The Minkowski values of
the southern part in  particular exhibit  drastic deviations from  the
reference Poisson samples.
\item
The  structural  difference between north  and  south is  also clearly
detectable  both in  the  nearest--neighbour distribution  and in  the
variance of galaxy  number counts.   The  clustering in the south   is
recognizably stronger than in the north.
\end{itemize}

A  similar anisotropy  in  the angular  distributions of IRAS galaxies
around the northern  and southern   galactic  poles has  already  been
reported by   {}\scite{rowan:sourcecount}    and was   confirmed    by
{}\scite{lonsdale:galaxyevolution}.  However, the majority of previous
IRAS     catalogue      studies   (e.g.\    {}\pcite{bouchet:moments},
{}\pcite{fisher:clusteringI},            {}\pcite{kaiser:large-sacle},
{}\pcite{oliver:largeiras},             {}\pcite{protogeros:topology},
{}\pcite{yess:percolation}) focused attention  on the  complete sample
without addressing the distinction of its constituent parts.

There is no reason to assume a distinguished position of the Milky Way
galaxy;  we   therefore  conclude  that  fluctuations  in  the  global
morphological  characteristics of  the     IRAS sample   extend   over
length--scales of 200\hMpc, at least. These fluctuations may originate
from dynamical correlations   in the matter distribution   which arise
during cosmic evolution.

Fluctuations occuring  on scales up  to 200\hMpc, at least, imply that
cosmological simulations which {\em  enforce} homogeneity on the scale
of a few hundreds of \hMpc\ and suppress fluctuations on larger scales
by   using   periodic   boundary   conditions   cannot reproduce   the
large--scale fluctuations  indicated by the   present analysis of  the
1.2~Jy catalogue. This assertion  is confirmed by our  comparison with
IRAS mock catalogues drawn from simulations of 256\hMpc\ box--length.

Recently  {}\scite{davis:homogeneous}   used  the  flux limited 1.2~Jy
galaxy catalogue without volume  limitation and looked at  galaxies in
different  redshift  intervals, projected onto   the  sphere. He found
strong anisotropies   in the number  density in  the nearby region but
less in the  deeper regions and claims  to see a cross--over towards a
random  homogeneous distribution   on scales  well  below the   sample
size. However, our results underline   that the visual impression   of
homogeneity on larger scales,  especially if based essentially on  the
number density, is indeed deceptive and misleading.

During  the last   few years,  {}\scite{pietronero:fractal},  see also
{}\scite{labini:frequently}     or   {}\scite{coleman:fractal},   have
advanced an   interpretation of galaxy  catalogue  data in terms  of a
fractal support  of the galaxy   distribution. By  its nature, a  pure
fractal indeed  exhibits fluctuations  in $N$--point distributions  on
all scales.   Our  results   neither  support nor     contradict  this
interpretation, since the Minkowski   functionals, as employed in  the
present  paper,    are  global  measures  and   are   not  designed to
discriminate local structures of spatial patterns.

It will be highly  interesting to repeat  the present analysis for the
upcoming   PSCz  catalogue,     and     for  an   optically   selected
survey. Forthcoming work will   also  focus on  the analysis   of mock
catalogues and cosmic  variance in  large ($\gtrsim1h^{-1}\text{Gpc}$)
simulations.

\section*{Acknowledgements}

We thank Thomas Boller, Stephanie C\^{o}t\'{e}, Luiz da Costa, Avishai
Dekel, Vicent  Mart\'{\i}nez, Maria  Jesus Pons--Border\'{\i}a, Gustav
Tammann,   Roberto  Trasarti--Battistoni,  and  Simon~D.M.  White  for
stimulating  discussions.  We are grateful   to Avishai Dekel and Yair
Sigad for kindly  providing  the simulation.   MK and   TB acknowledge
support from    the  {\em   Sonderforschungsbereich  SFB  375    f\"ur
Astroteilchenphysik  der    Deutschen Forschungsgemeinschaft} and from
Acciones Integradas during their  stay in Val\`{e}ncia where parts  of
this work were prepared.

\section*{}
The software for calculating the  Minkowski functionals, and also  the
other  measures  employed, can  be  obtained by  sending email to {\tt
buchert@stat.physik.uni-muenchen.de}.

\appendix

\section{Boundary corrections}
\label{appendix:boundaries}

This  appendix    summarizes boundary  corrected  estimators   for the
Minkowski functionals $M_\mu$, the nearest neighbour distributions $F$
and $G$ and for  the fluctuations of counts  in cells $\sigma^2$.  All
estimators are unbiased only for a statistically ``fair'' sample and a
stationary point process.  However, we use identical window geometries
for the northern and the southern  parts, so we  may still compare the
two   parts without making   these  assumptions. Moreover the observed
discrepancies can be used to falsify at least one of them.

\subsection{Boundary correction for Minkowski functionals}

Integral geometry  offers a concise  way of dealing with contributions
of     the   sample         geometry    {}\cite{mecke:euler}.       In
{}\scite{schmalzing:minkowski}  we   discuss  a  method for completely
removing boundary effects.

\begin{figure}
\begin{center} 
\epsfxsize=8cm
\begin{minipage}{\epsfxsize}\epsffile{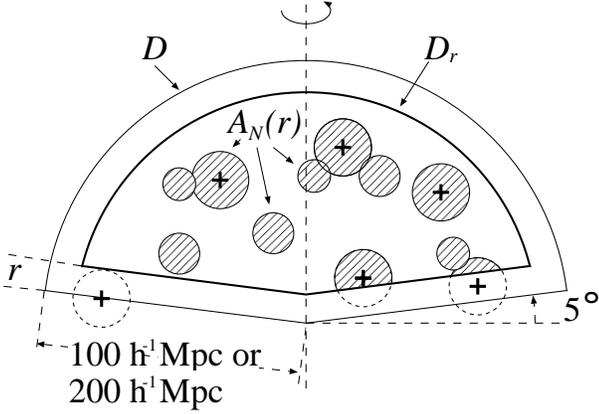}\end{minipage} 
\end{center}
\caption{\label{fig:window}   A  cut through   the union  set of balls
$\cA_N(r)$, within the shrunken window $D_r$.}
\end{figure}

Let $D$ be the window, i.e.\ the sample geometry through which we look
at the  galaxies. First of all,  we must shrink the  window $D$ by the
radius $r$, otherwise we  would miss contributions from balls situated
outside the sample (as  shown in Fig.~\ref{fig:window}). The  shrunken
window is $D_r:=D\uminus\cB_r$.

We  calculate the  Minkowski functionals  $M_\mu(D_r)$  of the  window
itself, and the Minkowski functionals $M_\mu(\cA_N(r)\cap D_r)$ of the
intersection of   the window  with     the union  of all  balls   (see
Fig.~(\ref{fig:window})).  From these numerically measured values, the
volume  densities $m_\mu(\cA_r)$ of  the  Minkowski functionals of the
union set without     the   window are  obtained   by   the  recursive
formula\footnote{We use the convention  $\textstyle\sum_{i=0}^{-1}a_i=
0$ to include the case $\mu=0$.}
\begin{multline}\label{eq:deconv}
m_\mu(\cA_N(r))=\\
\frac{M_\mu(\cA_N(r)\cap{D_r})}{M_0(D_r)}-
\sum_{\nu=0}^{\mu-1}\binom{\mu}{\nu}m_\nu(\cA_N(r))
\frac{M_{\mu-\nu}(D_r)}{M_0(D_r)}.
\end{multline}
Equation~(\ref{eq:deconv}) is just    a linear transformation  of  the
functionals $M_\mu(\cA_N(r)\cap  D_r)$  calculated for one realization
of   the point  process.  Hadwiger's Theorem~\cite{hadwiger:vorlesung}
asserts that the $m_\mu(\cA_N(r))$ are a complete set of morphological
descriptors equivalent to $M_\mu(\cA_N(r)\cap D_r)$. What is more, the
mean  values of   $m_\mu(\cA_N(r))$  are  analytically known  for  the
Poisson process,  and  can be  directly  compared  to the results   in
Eq.~(\ref{eq:Poisson}).  Finally, for a  fair  sample of a  stationary
point process  they are {\em unbiased  estimators} of volume densities
of the Minkowski functionals {}\cite{fava:plate,fava:random}.

\subsection{Boundary correction for nearest neighbour statistics}

{}\scite{ripley:spatial} gives   several  estimators for $F$   and $G$
applicable to spatial point processes with  boundaries. In our work we
exclusively    use the minus   estimators  (also called reduced sample
estimators) $\widehat{F}$ and $\widehat{G}$.

The definition uses the distance $\rho(\by,X)$ of a point $\by$ to the
nearest point of the pattern $X=\{\bx_i\}_{i=1}^N$,
\begin{equation}
\rho(\by,X):=\min(||\bx_i-\by||~|~
\bx_i{\in}X\mbox{ and }\bx_i\ne\by), 
\end{equation}
and the distance $\rho(\by,D)$ of a point $\by$ to the
boundary of the window $D$,
\begin{equation}
\rho(\by,D):=\min(||\bz-\by||~|~\bz\not\in{D}).
\end{equation}
As  explained  in   Sect.~\ref{sec:nextneighbour},  $F(r)$ gives   the
probability of finding a point  of the pattern  $X$ within a  distance
$r$ around a randomly selected point $\by$.  Therefore we estimate the
proportion of  points $\by\in{D}$ with $\rho(\by,X)\leq{r}$.  In order
to  obtain an unbiased  estimator, $\by$ has to  keep a distance of at
least  $r$   from the  window's   boundary, i.e.\  $r\leq\rho(\by,D)$.
Therefore, let $\{\by_j\}_{j=1}^M$  be $M$ randomly distributed points
$\by_j\in{D_r}$   inside   the    shrunken     window   $D_r$.    Then
$\widehat{F}(r)$ is defined by
\begin{equation}
\label{eq:estimator_F}
\widehat{F}(r) = \frac
{\#\{\by_j~|~\rho(\by_j,X)\leq{r})\}}
{M},
\end{equation}
where $\#\{\ldots\}$  gives     the number  of  points    in  the  set
$\{\ldots\}$.

Note   that Eq.~(\ref{eq:estimator_F}) is  nothing   but a Monte Carlo
integration of  the volume  $\cA_N(r)\cap    D_r$. This is   a  direct
consequence of the equality $F_0(r)=m_0(r)$ mentioned earlier.

To    estimate $G(r)$  we  also    use   a  reduced sample   estimator
{}\cite{ripley:spatial}.   Since  $G(r)$   gives   the probability  of
finding another point of the pattern $X$  within a distance $r$ around
a  point $\bx\in{X}$,  we simply  replace  the  random sample used  in
Eq.~(\ref{eq:estimator_F}) by the pattern   $X$.  The minus  estimator
$\widehat{G}(r)$ is then defined by
\begin{equation}
\widehat{G}(r) = \frac
{\#\{\bx_j~|~\rho(\bx_j,X)\leq{r}\leq\rho(\bx_j,D)\}}
{\#\{\bx_j~|~r\leq\rho(\bx_j,D)\}}.
\end{equation}

\subsection{Boundary correction for galaxy count fluctuations}

For  estimating $\sigma^2(r)$ as defined in Eq.~(\ref{eq:sigma-def1}),
we randomly throw  two million  balls  $\cB_r$,  count the number   of
points in each and calculate  the variance.  In order  to take care of
the boundaries we only choose balls lying completely within the window
$D$. This means  their centres have to  fall into  the shrunken window
$D_r$.

\section{$\sigma^2$ for a fractal, and dependence on 
\lowercase{$\overline{n}$}}
\label{appendix:fractal}

Figure~{}\ref{fig:sigma_comb}   reveals  differences   of   $\sigma^2$
between the 100\hMpc\ and the 200\hMpc\ samples.  Although some of the
results are   still consistent given   the relatively large errorbars,
another  explanation of the differences  is possible. In this appendix
we  calculate the dependence of $\sigma^2$  on the depth $R_0$ and the
number of points $N_0$ for a sample drawn  from a fractal distribution
of galaxies.

The generalization of $\sigma^2$ to point processes on fractal sets is
not straightforward, since the  mean number density $\overline{n}$  is
not       well--defined.     In     the      definition   given     in
Eq.~(\ref{eq:sigma-def1}) we have to replace the number of points in a
ball of radius $r$ by $N(r)$ instead of $\overline{n}V(\cB_r)$, so
\begin{equation}\label{eq:sigma-def2}
\sigma^2(r) := 
\frac{\left\langle(N_i-N(r))^2\right\rangle-N(r)}{N^2(r)}
\end{equation}
For a fractal of correlation dimension $D$ this quantity depends on
the radius through
\begin{equation}\label{eq:Nr-def}
N(r) = \left(\frac{N_0}{R_0^D}\right) r^D,
\end{equation}
where the   constant  of normalization $N_0/R_0^D$    is  given for  a
spherical   sample  of radius $R_0$  containing   $N_0$ galaxies.  The
generalization    to cones and   an  application  to  number counts is
discussed  in {}\scite{labini:numbercount}. With Eq.~(\ref{eq:Nr-def})
plugged into the definition in Eq.~(\ref{eq:sigma-def2}) we obtain
\begin{equation}
\sigma^2(r)= 
\frac{\left\langle{N_i^2}\right\rangle}{N_0^2}
\left(\frac{R_0}{r}\right)^{2D} 
-\frac{1}{N_0}\left(\frac{R_0}{r}\right)^{D} -1.
\end{equation}
Obviously $\sigma^2$ depends on $N_0$ and $R_0$ nontrivially if $D<3$.
For   a   stationary   process  with    $D=3$  the    average  density
$\overline{n}:=N_0/V_0=N_0/(\frac{4\pi}{3}R_0^3)$  is  independent  of
the sample size and we have
\begin{equation}\label{eq:sigma-n}
\sigma^2(r) = \left\langle{N_i^2}\right\rangle 
(\overline{n}\textstyle\frac{4\pi}{3}r^3)^{-2} -
(\overline{n}\textstyle\frac{4\pi}{3}r^3)^{-1} -1
\end{equation}
as  a  function of the  mean   number density  $\overline{n}$.   Since
$\langle\cdot\rangle$    is   defined     as     a   volume    average,
$\left\langle{N_i^2}\right\rangle$ determined for one realization  may
differ from an average over several realizations, i.e.\ universes.  If
one   assumes   ergodicity  of    the  point  distribution   (implying
homogeneity),  the  volume  average  converges   towards the  ensemble
average.  In  this case  we  get  for  example for  a Poisson  process
$\left\langle{N_i^2}\right\rangle                                    =
(\overline{n}\textstyle\frac{4\pi}{3}r^3)^{2}+
\overline{n}\textstyle\frac{4\pi}{3}r^3$ and hence $\sigma^2(r)=0$.


\providecommand{\bysame}{\leavevmode\hbox to3em{\hrulefill}\thinspace}


\begin{thebibliography}{{Rowan-Robinson \bgroup et al.\egroup }{1986}}

\bibitem[\protect\citefmt{Bouchet \bgroup et al.\egroup
  }{1993}]{bouchet:moments}
Bouchet F.~R., Strauss M.~A., Davis M. et~al., 1993, ApJ 417, 36

\bibitem[\protect\citefmt{Coleman and Pietronero}{1992}]{coleman:fractal}
Coleman P.~H., Pietronero L., 1992, Physics Rep. 213, 311

\bibitem[\protect\citefmt{Davis}{1996}]{davis:homogeneous}
Davis M.: 1996, \emph{Is the {U}niverse homogeneous on large scales?}. In:
  \emph{Critical Dialogues in Cosmology}, Turok, N. (ed.), astro-ph/9610149

\bibitem[\protect\citefmt{Fava and Santal{\'o}}{1978}]{fava:plate}
Fava N.~A., Santal{\'o} L.~A., 1978, J.\ Appl.\ Prob. 15, 494

\bibitem[\protect\citefmt{Fava and Santal{\'o}}{1979}]{fava:random}
Fava N.~A., Santal{\'o} L.~A., 1979, Z.\ Wahrscheinlichkeitstheorie verw.\
  Gebiete 50, 85

\bibitem[\protect\citefmt{{Fisher} \bgroup et al.\egroup
  }{1994}]{fisher:clusteringI}
{Fisher} K.~B., {Davis} M., {Strauss} M.~A. et~al., 1994, MNRAS 266, 50

\bibitem[\protect\citefmt{Fisher \bgroup et al.\egroup
  }{1995}]{fisher:irasdata}
Fisher K.~B., Huchra J.~P., Strauss M.~A. et~al., 1995, ApJS 100, 69

\bibitem[\protect\citefmt{Hadwiger}{1957}]{hadwiger:vorlesung}
Hadwiger H., 1957, \emph{Vorlesungen {\"u}ber {I}nhalt, {O}berfl{\"a}che und
  {I}soperimetrie}, Springer Verlag, Berlin

\bibitem[\protect\citefmt{Kaiser \bgroup et al.\egroup
  }{1991}]{kaiser:large-sacle}
Kaiser N., Efstathiou G., Ellis R. et~al., 1991, MNRAS 252, 1

\bibitem[\protect\citefmt{Kerscher \bgroup et al.\egroup
  }{1996}]{kerscher:significance}
Kerscher M., Schmalzing J., Buchert T., Wagner H.: 1996, \emph{The
  significance of the fluctuations in the {IRAS} 1.2 {J}y galaxy catalogue}.
  In: \emph{Proc.\ $2^{\rm nd}$ SFB workshop on {\em Astro--particle physics}
  Ringberg 1996, Report SFB375/P002} (Ringberg, Tegernsee), Bender, R.,
  Buchert, T., Schneider, P. (eds.), p.~83

\bibitem[\protect\citefmt{Kerscher \bgroup et al.\egroup
  }{1997}]{kerscher:abell}
Kerscher M., Schmalzing J., Retzlaff J. et~al., 1997, MNRAS 284, 73

\bibitem[\protect\citefmt{Kolatt \bgroup et al.\egroup
  }{1996}]{kolatt:simulating}
Kolatt T., Dekel A., Ganon G., Willick J.~A., 1996, ApJ 458, 419

\bibitem[\protect\citefmt{Lonsdale \bgroup et al.\egroup
  }{1990}]{lonsdale:galaxyevolution}
Lonsdale C.~J., Hacking P.~B., Conrow T.~P., Rowan-Robinson M., 1990, ApJ
  358, 60

\bibitem[\protect\citefmt{Mann \bgroup et al.\egroup }{1996}]{mann:warmcool}
Mann R.~G., Saunders W., Taylor A.~N., 1996, MNRAS 279, 636

\bibitem[\protect\citefmt{Mecke and Wagner}{1991}]{mecke:euler}
Mecke K.~R., Wagner H., 1991, J.\ Stat.\ Phys. 64, 843

\bibitem[\protect\citefmt{Mecke \bgroup et al.\egroup }{1994}]{mecke:robust}
Mecke K.~R., Buchert T., Wagner H., 1994, A\&A 288, 697

\bibitem[\protect\citefmt{Oliver \bgroup et al.\egroup
  }{1996}]{oliver:largeiras}
Oliver S.~J., Rowan-Robinson M., Broadhurst T.~J. et~al., 1996, MNRAS 280,
  673

\bibitem[\protect\citefmt{Peebles}{1993}]{peebles:principles}
Peebles P.~J.~E., 1993, \emph{Principles of physical cosmology}, Princeton
  University Press, Princeton, New Jersey

\bibitem[\protect\citefmt{Pietronero \bgroup et al.\egroup
  }{1996}]{pietronero:fractal}
Pietronero L., Montuori M., {Sylos Labini} F.: 1996, \emph{On the fractal
  structure of the visible {U}niverse}. In: \emph{Critical Dialogues in
  Cosmology}, Turok, N. (ed.), astro-ph/9611197

\bibitem[\protect\citefmt{Protogeros and Weinberg}{1997}]{protogeros:topology}
Protogeros Z. A.~M., Weinberg D.~H., 1997, ApJ 489, 457

\bibitem[\protect\citefmt{Ripley}{1988}]{ripley:spatial}
Ripley B.~D., 1988, \emph{Statistical inference for spatial processes},
  Cambridge University Press, Cambridge

\bibitem[\protect\citefmt{Rowan-Robinson \bgroup et al.\egroup
  }{1986}]{rowan:sourcecount}
Rowan-Robinson M., Walker D., Chester T. et~al., 1986, MNRAS 219, 273

\bibitem[\protect\citefmt{Schmalzing \bgroup et al.\egroup
  }{1996}]{schmalzing:minkowski}
Schmalzing J., Kerscher M., Buchert T.: 1996, \emph{{M}inkowski functionals
  in cosmology}. In: \emph{Proceedings of the international school of physics
  Enrico Fermi. Course CXXXII: Dark matter in the {U}niverse} (Varenna sul Lago
  di Como), Bonometto, S., Primack, J., Provenzale, A. (eds.), Societ{\`a}
  Italiana di Fisica, p.~281

\bibitem[\protect\citefmt{{Sylos Labini} \bgroup et al.\egroup
  }{1996}]{labini:numbercount}
{Sylos Labini} F., Gabrielli A., Montuori M., Pietronero L., 1996, Physica
  A 226, 195

\bibitem[\protect\citefmt{{Sylos Labini} \bgroup et al.\egroup
  }{1997}]{labini:frequently}
{Sylos Labini} F., Pietronero L., Montuori M.: 1997, \emph{Frequently asked
  questions about fractals}. In: \emph{Proc.\ $2^{\rm nd}$ SFB workshop on {\em
  Astro--particle physics} Ringberg 1996, Report SFB375/P002} (Ringberg,
  Tegernsee), Bender, R., Buchert, T., Schneider, P. (eds.), p.~109

\bibitem[\protect\citefmt{{van Lieshout} and Baddeley}{1996}]{vanlieshout:j}
{van Lieshout} M.~N.~M., Baddeley A.~J., 1996, Statist.\ Neerlandica 50,
  344

\bibitem[\protect\citefmt{White}{1979}]{white:hierarchy}
White S. D.~M., 1979, MNRAS 186, 145

\bibitem[\protect\citefmt{Yess \bgroup et al.\egroup }{1997}]{yess:percolation}
Yess C., Shandarin S.~F., Fisher K.~B., 1997, ApJ 474, 553

\end{thebibliography}
\end{document}